\documentclass[groupedaddress]{revtex4}
\usepackage{graphicx} 
\usepackage{amssymb} 
\usepackage{amsmath} 
\usepackage{epsfig} 
\usepackage{xspace} 
\usepackage{color} 
\usepackage[utf8]{inputenc}
\usepackage{ulem}
\usepackage{hyperref}
\usepackage{url}
\definecolor{MyDarkBlue}{rgb}{0,0.0,0.7}
\hypersetup{pdfborder={0 0 0},colorlinks,breaklinks=true,
  urlcolor={MyDarkBlue},citecolor={MyDarkBlue},linkcolor={MyDarkBlue}
}

\newcommand{\C}{\mathbb{C}} 
 
\newcommand{\R}{\mathbb{R}}

\newcommand{\1}{\mbox{1\hspace{-.65ex}I}}

\newcommand{\cL}{{\cal L}} 
 
\newcommand{\cM}{{\cal M}} 
\newcommand{\cO}{{\mathcal O}}

\newcommand{\bra}{\langle} 
\newcommand{\ket}{\rangle}

\newcommand{\bear}{\begin{eqnarray}} 
\newcommand{\eear}{\end{eqnarray}}

\renewcommand{\Re}{{\rm Re}} 
\renewcommand{\Im}{{\rm Im}}

\def\be#1\ee{\begin{equation}#1\end{equation}} 
\def\bea#1\eea{\begin{align}#1\end{align}}

\newcommand{\tr}{\textrm{Tr}\,}

\renewcommand{\vec}[1]{{\bf #1}}

\begin{document}


\title{Controlling Complex Langevin simulations of lattice models \\by 
boundary term analysis} \author{M. Scherzer$^{1}$} \author{E. 
Seiler$^{2}$} \author{D. Sexty$^{3,4}$} \author{I.-O. Stamatescu$^{1}$}

\affiliation{$^1$ {\it Institut f\"ur Theoretische Physik, Universit\"at 
Heidelberg, Philosophenweg 16, 69120 Heidelberg, Germany}} 
\affiliation{$^2$ {\it Max-Planck-Institut f\"ur Physik 
(Werner-Heisenberg-Institut), F{\"o}hringer Ring 6, 80805 M{\"u}nchen, 
Germany}} \affiliation{ $^3$ {\it Department of Physics, Wuppertal 
University, Gau{\ss}str. 20, D-42119 Wuppertal, Germany}} \affiliation{ 
$^4$ {\it J\"ulich Supercomputing Centre, Forschungszentrum J\"ulich, 
D-52425 J\"ulich, Germany}}

\begin{abstract} 
\noindent 
One reason for the well known fact that the Complex Langevin (CL) method 
sometimes fails to converge or converges to the wrong limit has been 
identified long ago: it is insufficient decay of the probability density 
either near infinity or near poles of the drift, leading to boundary terms 
that spoil the formal argument for correctness. To gain a deeper 
understanding of this phenomenon, in a previous paper 
\cite{boundaryterms1} we have studied the emergence of such boundary 
terms thoroughly in a simple model, where analytic results can be compared 
with numerics. Here we continue this type of analysis for more physically 
interesting models, focusing on the boundaries at infinity. We start with 
abelian and non-abelian one-plaquette models, then we proceed to a 
Polyakov chain model and finally to high density QCD (HDQCD) and the 3D XY 
model.
We show that the direct estimation of the systematic error of the CL method
 using boundary terms is in principle possible.
\end{abstract} 

\maketitle

\section{Introduction}

Complex Langevin simulations are a very general method which can in 
principle be applied to any model with complex action, allowing an
analytic continuation into the complexification of the original configuration
space. The setup is straightforward and needs 
no preliminary steps, such as model dependent design or approximations. 
These features motivate the work to ensure the
  reliability of Complex Langevin simulations,
since the resulting stochastic processes in the complexified configuration
space require care due to their mathematical subtleties. 

This paper extends to realistic lattice models the study of boundary terms 
\cite{boundaryterms1} which occur in some Complex Langevin (CL) 
simulations and have the undesired effect of spoiling correctness. We thereby
aim at the estimation of possible systematic errors and
correction of the results.

We briefly collect some necessary definitions to make this paper 
self-contained. For more details we refer to \cite{boundaryterms1} as 
well as to earlier papers such as \cite{Aarts:2009uq,Aarts:2011ax,Seiler:2017wvd}.

The complex Langevin (CL) process defines a time dependent probability 
density $P(t)$ on the complexification $\cM_c$ of the original 
configuration space $\cM$, so we sometimes write it as $P(\vec x,\vec 
y;t)$ where $\vec x$ stands for the real and $\vec y$ for the imaginary 
part of the configuration variables.  For notational simplicity we assume 
that $\cM$ and $\cM_c$ are flat, with coordinates $\vec x$ and $\vec 
x+i\vec y $, respectively; we will indicate the necessary changes for the 
non-flat case later.

 $P(\vec x, \vec y;t)$ obeys the Fokker-Planck equation 
(FPE) \bea \partial_t P(\vec x,\vec y,t) = L^T P, \qquad L^T= \nabla_{x} 
\cdot (\nabla_{x} -\vec K_{x}) - \nabla_y \cdot \vec K_{y}\, 
\label{eq:real_FP} \eea with
\be \vec K_x(\vec x, \vec y) = -\Re\nabla S\,,\quad \vec K_y(\vec x,\vec y)= 
-\Im \nabla S\,, \ee
where $S$ is the action entering the
integration measure $\rho=e^{-S}$ in the partition function; (\ref{eq:real_FP}) 
determines the time dependent expectation values of holomorphic 
observables $\cO$ via \be \bra \cO \ket_{P(t)}=\int P(\vec x,\vec 
y;t)\cO(\vec x+ i\vec y) d^Nx d^Ny\,. \label{pevol} \ee This is to be 
compared with the `correct evolution'
\be \bra \cO \ket_{\rho(t)}=\int 
\rho(\vec x;t)\cO(\vec x)d^Nx\, \label{rhoevol} \ee
computed using an 
evolution of the complex density $\rho(t)$ on the original real 
configuration space $\cM$ determined by the PDE (`complex FPE')
\be 
\partial_t \rho(\vec x;t)= L_c^T \rho(\vec x;t)\,, \qquad L_c^T = 
\nabla_x\cdot (\nabla_x -\vec K_x(\vec x)-i\vec K_y(\vec x))\,; \label{complex_FP} \ee
Correctness of the CL evolution means then equality of Eqs (\ref{pevol}) 
and (\ref{rhoevol}). Equality and hence correctness of the evolution 
depends on (1) equality at $t=0$, which can be easily arranged; (2) the 
absence of boundary terms both at infinity and near poles of the drift. 
For the models we study here poles are either absent or far away from the 
$P$ distribution and do not play a relevant role.

Correct convergence for $t\to\infty$ depends in addition on existence and 
uniqueness (independence of the initial conditions) of the limit \be 
\lim_{t\to\infty} \bra \cO \ket_{\rho(t)}\,, \ee which depends on the 
spectrum of $L^T_c$ being located in the left half of the complex 
plane with a simple eigenvalue at the origin; the latter property is 
closely related to ergodicity of the CL process; this is a problem for stochastic processes in general.

The study of possible boundary terms uses a function $F_{\cO}$ 
interpolating between the two evolutions \be F_\cO(t,\tau)\equiv \int 
P(\vec x,\vec y;t-\tau) \cO(\vec x+i\vec y;\tau)d^Nxd^Ny\,. \label{fttau} 
\ee $F_{\cO}$ satisfies \be F_{\cO}(t,0)=\bra \cO \ket_{P(t)}\,,\quad 
F_{\cO}(t,t)=\bra \cO \ket_{\rho(t)}\,, \ee such that correctness of the 
evolution is guaranteed if \be \frac{\partial}{\partial \tau} 
F_\cO(t,\tau) = 0\,. \label{interpol} \ee.

\section{Two versions of boundary terms}

\subsection{Boundary term as integral over the surface}

As discussed in \cite{boundaryterms1}, the left hand side of 
(\ref{interpol}) is really a boundary term. We also found there that 
typically this derivative is maximal at $\tau=0$, so we focus on \be 
\partial_\tau F_{\cO}(t,\tau)|_{\tau=0}.  \ee We 
rewrite this as an explicit boundary term, still assuming $\cM$ and 
$\cM_c$ as flat.
Suppressing the configuration arguments $\vec x, \vec y$ and introducing a cutoff $Y$ on the imaginary part $\vec{y}$ in
(\ref{fttau}), we define
\be F_\cO(Y;t,\tau)\equiv \int_{|\vec{y}|\leq Y} 
P(\vec x,\vec y;t-\tau) \cO(\vec x+i\vec y;\tau)d^Nxd^Ny\,. \label{fttauYcut} 
\ee
and with  
(\ref{eq:real_FP},\ref{complex_FP}) we get the boundary term
\be \label{firstbtermeq}
\partial_\tau F_{\cO}(Y;t,\tau)|_{\tau=0}\equiv B(Y,t) = -\int_{|\vec y|\leq Y} \left(L^T P(t) \right) \cO(0)  d^Nx d^Ny
+ \int_{|\vec y|\leq Y} P(t) (L_c \cO(0)) d^Nx d^Ny\,. \ee Here we assumed 
  that the $\vec x$ integration is unproblematic because of periodicity or 
  fast decay so that the $\nabla_\vec x^2$ terms cancel by partial 
  integration, otherwise there would also
    be some $\vec x$ boundary terms, see e.g. in \cite{Aarts:2013uza}
    where the stationary distribution was found to be $P(x,y) \sim (x^2+y^2)^{-3/2} $.
    In this case one could calculate $x$ boundary terms by introducing a
    cutoff also on the $x$ coordinates in eq. (\ref{firstbtermeq}).

After some trivial manipulations (see \cite{boundaryterms1}), involving 
(assumed unproblematic) integration by parts in $\vec x$ and the Cauchy-Riemann 
equations,
\bea B(Y,t) = \int_{|\vec y|\leq Y} \nabla_{y} \left( \vec K_{y}\cO(0)  
P(t) \right) d^Nxd^Ny \eea
with the derivatives acting on everything to the right, 
so we are integrating a divergence. This is equal to the surface 
integral \bea B(Y,t)= \int_{|\vec y|=Y}\vec n\cdot \vec K_{y} P(t) \cO(0)  
d^Nx\, dS\,, \label{surface} \eea where $\vec n$ is the outer normal to 
the surface $|\vec y|=Y$ and $dS$ the surface element on ${|\vec y|=Y}$. 
Of course it is not necessary to choose the cutoff $Y$ in the form $|\vec 
y|\leq Y$ as we have done here; it is only necessary that the family of cutoffs 
restricts $\vec y$ to compact sets which exhaust the full space as we send 
$Y\to\infty$. Finally we take the limit $t \rightarrow \infty$ to extract
  $B(Y)$ in the stationary state.

\subsection{Boundary term as a volume integral}

To explain the principle we assume again that the configuration space 
$\cM$ is flat. Later we will see what has to be changed for the more 
interesting case of $\cM$ being a compact group manifold.

Proceeding as in \cite{boundaryterms1} we determine $B$ via a limiting 
procedure\bea B(Y)= \lim _ {t \rightarrow \infty} \partial_\tau F_{\cO}(Y;t,\tau=0)\,, \label{boundterm} \eea with $Y$ as 
before and still assuming that the real directions are either compact or 
have sufficient falloff to avoid any boundary terms there. $Y$ will be 
sent to $\infty$ later (cf. \cite{boundaryterms1}).

Now we process the term as follows: evaluating (\ref{boundterm}) we find 
\be \partial_\tau F_{\cO}(Y;t,\tau=0) = - \int_{|\vec y|\leq Y} (L^T P(\vec x, 
\vec y;t)) O(\vec x+i\vec y) d^Nx d^Ny +\int_{|\vec y|\leq Y} P(\vec x,\vec y,t) L_c \cO(\vec 
x+i\vec y) d^Nx d^Ny\,. \ee The $t\rightarrow \infty$ limit of the first term 
is zero as the process reaches equilibrium. The second term can be 
nonzero, spoiling correctness. So we have to study
\bea B(Y) = 
\int_{|\vec y|\leq Y} P(x,y,t=\infty) L_c \cO(x+i y) dx dy \label{bound_cc} 
\eea
Vanishing of $B(\infty)$ is just the old `consistency 
condition' or `convergence condition' (CC), discussed in 
\cite{Aarts:2011ax}, which signals stationarity.

We now describe briefly the changes to be made in the case where the 
configuration space $\cM$ is a compact group. Without loss of generality 
we may think of $\cM$ as a space of unitary matrices and $\cM_c$ a space 
of complex invertible matrices. Each matrix $M\in\cM_c$ has a polar 
decomposition \be M=RU \ee with $U$ unitary and $R=\sqrt{M^\dagger M}$ 
positive. We introduce a `unitarity norm' $UN$ (not a norm in the 
mathematical sense) to measure the distance of a $M\in\cM$ from the 
unitary subspace; a simple choice uses
\be 
n(M)=\tr(M^\dagger M-\1)^2\,, 
\label{unorm} 
\ee 
and defines $UN$ for a lattice model as $n(M)$ divided by the number of links or the maximum of $n(M)$ over the links.

The boundary term is given by
\be B(Y) = \int_{UN\leq Y} P(M;t=\infty) L_c \cO(M) dM \label{bound_ccgroup} 
\ee
where $dM$ is Haar measure on $\cM$. Also the operator $L_c$ has a 
slightly different form (see \cite{gaugecooling}): \be L_c=\sum_i (D_i+K_i) D_i\,, 
\ee where the operators $D_i$ are invariant vector fields on $\cM$, acting 
as derivations in the directions of a basis of the Lie algebra of $\cM$.

\section{Numerical Results}

\subsection{U(1) one plaquette model}

We revisit the one plaquette model with regularization
\bea \label{u1oneplaq}
S=i \beta \cos(x) + {s\over 2} x^2 \eea
with $x \in \R$ as investigated in 
\cite{boundaryterms1}. We calculate the boundary terms for the 
observable $\cO=\exp(i x) $ using the volume integral formulation, see 
\cite{boundaryterms1} for surface integration.
The boundary terms in this case arise 
by integrating
\bea L_c \exp(ikz) = i k ( i k + i \beta \sin(z) - s z) 
e^{i k z} \eea
with the measure $P(x,y;t=\infty)$ using the cutoff $Y$, as written in eq.(\ref{bound_cc}).

\begin{figure} 
\includegraphics[width=0.48\textwidth]{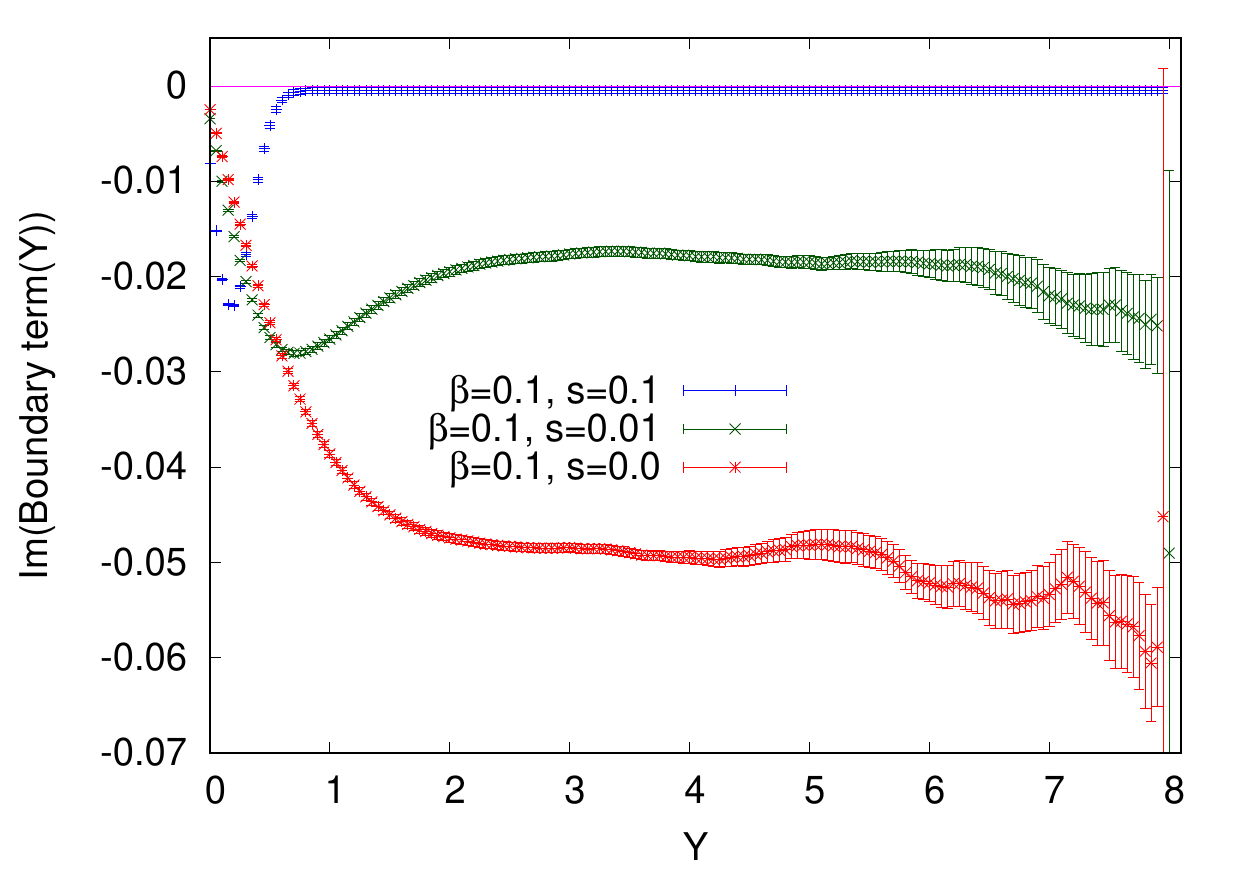} 
\caption{The imaginary part of the boundary term for the observable
  $e^{i x}$ is shown as a function of $Y$ for $\beta=0.1$ and
  several $s$ values in the $U(1)$ one-plaquette model.} \label{fig:ycc1}
\end{figure}

At $s=0$ there 
is a boundary term persisting for $Y\to\infty$, as can be inferred from 
Fig.~\ref{fig:ycc1}. 
  At large $Y$ we have more and more 
fluctuations, so the result becomes submerged in the noise.
 Finally, for large enough $Y$, with our limited statistics and necessarily 
finite Langevin time, we get the same result as without cutoff, since no 
points outside the cutoff region are sampled by the CL process and
hence also not 
discarded. This corresponds to taking the limits in the opposite order, 
i.e. first $Y\to\infty$ and then $t\to\infty$. The CC, expressing 
equilibrium, have to be fulfilled in this limit, albeit with potentially 
very large fluctuations. This is seen in the last red and green data points.

For sufficiently large $s$ we see that the boundary term converges to a value
consistent with $0$ and, 
as found in \cite{boundaryterms1}, the CL simulation gives the correct 
results of the regularized model within errorbars.

 In Sec.~\ref{sec:estu1} we show that the systematic error of the CL result
  is directly related to the boundary term measured here, and it can be
  estimated using the CL simulation alone. 
  
\subsection{$SU(3)$ one plaquette model and Polyakov chain}

Now we investigate the holomorphic Polyakov chain for possible boundary 
terms. The chain is defined via \bea \rho&=\exp(-S) \nonumber \\ -S&=c_+ 
\tr\cL +c_- \tr\cL^{-1}\, , \eea where $c_{\pm}=\beta+\kappa 
\text{exp}(\pm \mu)$, $\cL$ is the Polyakov loop
\bea
\cL & =  U_0 U_1 \,\ldots U_{N-1} \\ \nonumber
\cL^{-1} &= U_{N-1}^{-1} U_{N-2}^{-1}\ldots U_0^{-1}\, ,
\eea
and the $U_i$ are $SU(3)$ matrices associated to the $N$ 
links, analytically continued in the CL process to $SL(3,\C)$. $S$ is a 
holomorphic action, hence deviations of the CL result should only come from boundary 
terms at infinity. The model has a gauge symmetry that makes all $N$ values 
equivalent, but it presents a good test-bed for simulation methods.

We simulate this model in two different ways.

\subsubsection{Gauge fixing at $N=1$}

Here we use the gauge symmetry to diagonalize the matrix $U$. Since for 
$U\in SL(3,\C)$ $\det\,U=1$, this means that there are only two 
degrees of freedom. A single link now reads \begin{equation} 
U=\text{diag}\left(e^{i\omega_1}, e^{i\omega_2}, 
e^{-i(\omega_1+\omega_2)}\right), \quad
|\omega_1|,|\omega_2|,|\omega_1+\omega_2| \le \pi
\label{link} \end{equation} and the action becomes \begin{equation} 
-S=c_+\left(e^{i\omega_1}+ e^{i\omega_2}+ 
e^{-i(\omega_1+\omega_2)}\right)+c_-\left(e^{-i\omega_1}+ e^{-i\omega_2}+ 
e^{i(\omega_1+\omega_2)}\right)\,. \end{equation} In addition one has to 
include the reduced Haar measure, which adds to the action the term 
\begin{equation}
 -S_\text{meas}= \text{ln} 
\left[\text{sin}^2\left(-\frac{2\omega_1+\omega_2}{2}\right) 
\text{sin}^2\left(\frac{\omega_1-\omega_2}{2}\right)\text{sin}^2 
\left(\frac{\omega_1+2\omega_2}{2}\right)\right]\,, \quad 
S_\text{tot}=S+S_\text{meas}\,. 
\end{equation} 
This term is not holomorphic and leads to poles in the drift; these are, 
however, located at the boundary of the domain specified in (\ref{link}) 
and therefore cannot lead to ergodicity problems; they also do not lead to 
boundary terms (cf. \cite{Aarts:2012ft}). This non-holomorphicity created by 
gauge fixing is innocuous.

The boundary term for $\bra\tr U\ket$  arises from the integrand \begin{align} 
 L_c \tr U &=\left(\nabla+\vec{K}\right)\nabla \tr U\nonumber\\
&=-\left(e^{i\omega_1}+2e^{-i(\omega_1+\omega_2)}+e^{i\omega_2}\right)
\nonumber\\
&+iK_{1}\left(e^{i\omega_1}-e^{-i(\omega_1+\omega_2)}\right)
+iK_{2}\left(e^{i\omega_2}-e^{-i(\omega_1+\omega_2)}\right). \end{align} 
with $ K_{i}=-\partial_{\omega_i}S_\text{tot}$, $i=1,2$.


The expression (\ref{surface}) for the boundary term can be used here 
straightforwardly. We calculate it for this model explicitly by defining a
surface on the complex manifold spanned by $\omega_1 $ and $\omega_2$ as 
the boundary of the compact domain $ Y \le Y_{cut}$ with 
$Y=\text{max}(|\text{Im}\omega_1|,|\text{Im}\omega_1|)$. The boundary term 
reads (dropping $t$ and $\tau$ dependence for briefness sake), defining 
$\vec x=(\Re\,\omega_1,\Re\,\omega_2)^T$ and $\vec y= 
(\Im\,\omega_1,\Im\,\omega_2)^T$

\begin{equation} \int \int\left[ \left(\vec{K}_y P(\vec{x},\vec{y}) 
\mathcal{O}(\vec{x}+i\vec{y})\right)\cdot\vec{n}\right]d\vec{x}dS_y 
\end{equation}
with the surface element $dS_y$.
This integral can be `measured' in the CL simulation. The measurement 
becomes harder with increasing $Y$ as the statistics deteriorates.

\begin{figure}[h] \centering 
\includegraphics[width=0.45\textwidth]{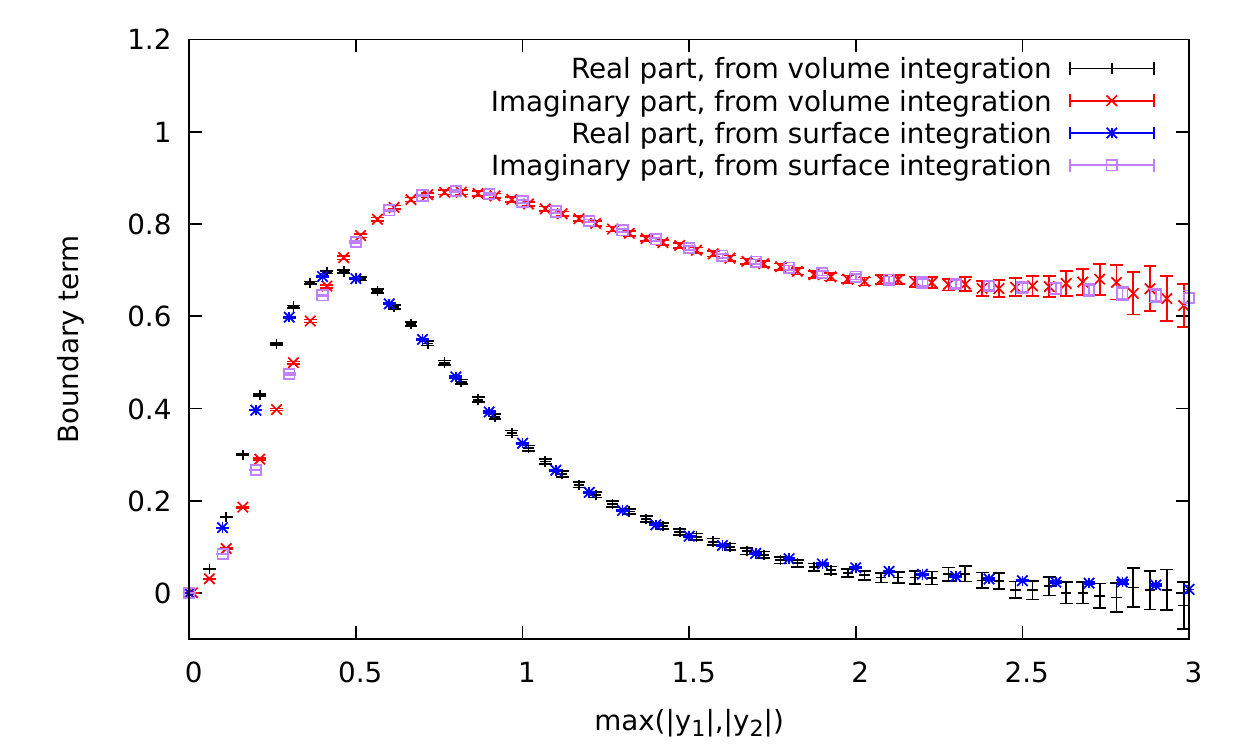} 
\includegraphics[width=0.45\textwidth]{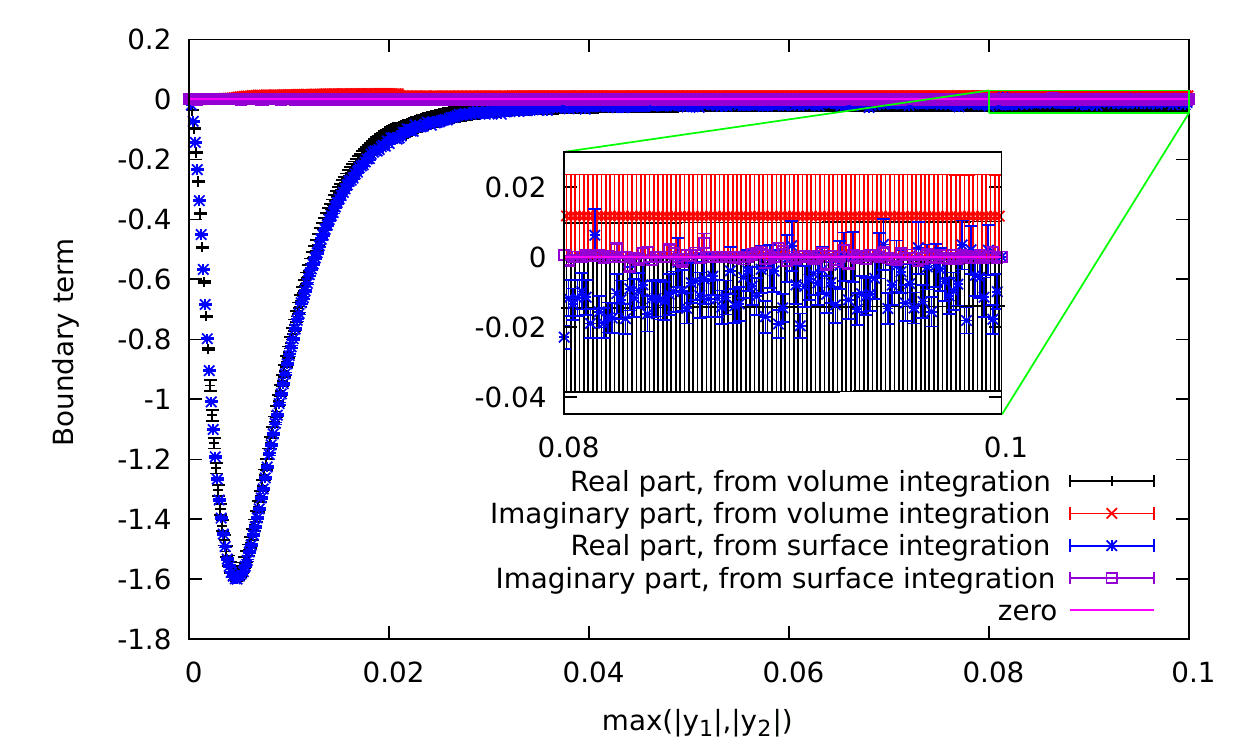} 
\caption{Comparison of boundary terms from volume and surface computation, 
showing agreement within the errors. Left: $\beta=i$, $\kappa=0=\mu$. 
Right: $\beta=2\,,\kappa=0.1\,,\mu=1$.} \label{fig:NI0} \end{figure}

We carried out a simulation for $\beta=i$, $\kappa=0=\mu$; the exact 
result for the Polyakov loop is $\bra \tr U \ket=-0.664+0.793i$, whereas 
the simulation yields $\bra \tr U \ket=-0.4809(6)+0.5968(5)i$, which is 
clearly not correct, i.e. boundary terms are to be expected. Fig. 
\ref{fig:NI0} left shows the boundary terms for this case, computed both 
in the surface and volume forms. We also did a run for 
$\beta=2\,,\kappa=0.1\,,\mu=1$ (see fig. \ref{fig:NI0}, right), where the 
simulation yields $\bra \tr U \ket=2.0955(13)$ which is consistent with 
the exact result $\bra \tr U \ket=2.0957$.

\subsubsection{Polyakov chain with $N>1$ and gauge cooling}

For $N>1$ we could use of course gauge fixing to reduce the model to 
$N=1$. It is more instructive, however, to leave all the link degrees of 
freedom and study the effect of having gauge degrees of freedom and
that of the gauge cooling \cite{gaugecooling} on the presence 
or absence of boundary terms; gauge cooling reduces the unitarity norm by 
non-compact gauge transformations. 

We use the volume form of the boundary terms in the following. The 
unitarity norm $UN$ used here is the average of $n(U_j)$ (see (\ref{unorm})) 
over the links. There are many sets of parameters for which CL without 
gauge cooling does not give correct results. As an illustration we choose 
$\beta=2.0$, $\kappa=0.1$, $\mu=1.0$, where the exact result is 
$\left<\tr\cL\right>=2.0957$. The boundary term integrand reads
\begin{align} 
L_c\tr \cL&=\sum_{j=0}^{N_t-1}\left(D_a^j+K_a^j\right) 
D_a^j\tr\cL \nonumber\\
&=-2N_t\frac{N^2-1}{N}\tr \cL+i\sum_{j=0}^{N_t-1}K_a^j
\text{Tr}\left(U_0\ldots\lambda_a U_j\ldots U_{N_t-1}\right)\,, 
\label{pol.bound}
\end{align} 
where the index $a$ refers to the standard basis of the $SU(3)$ Lie 
algebra given by the Gell-Mann matrices $\lambda_a\,,i=1,\ldots 8$ and $j$ 
numbers the link in the chain.

\begin{figure}[h] \centering 
\includegraphics[width=0.45\textwidth]{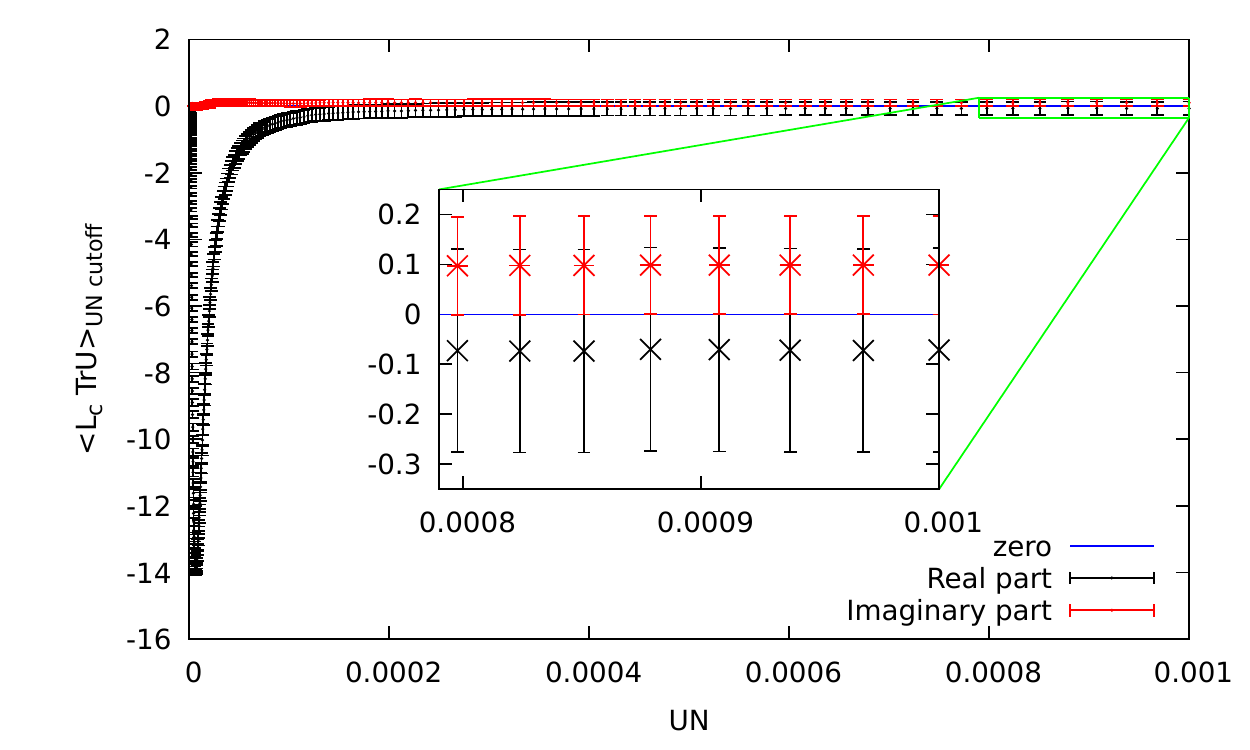} 
\includegraphics[width=0.45\textwidth]{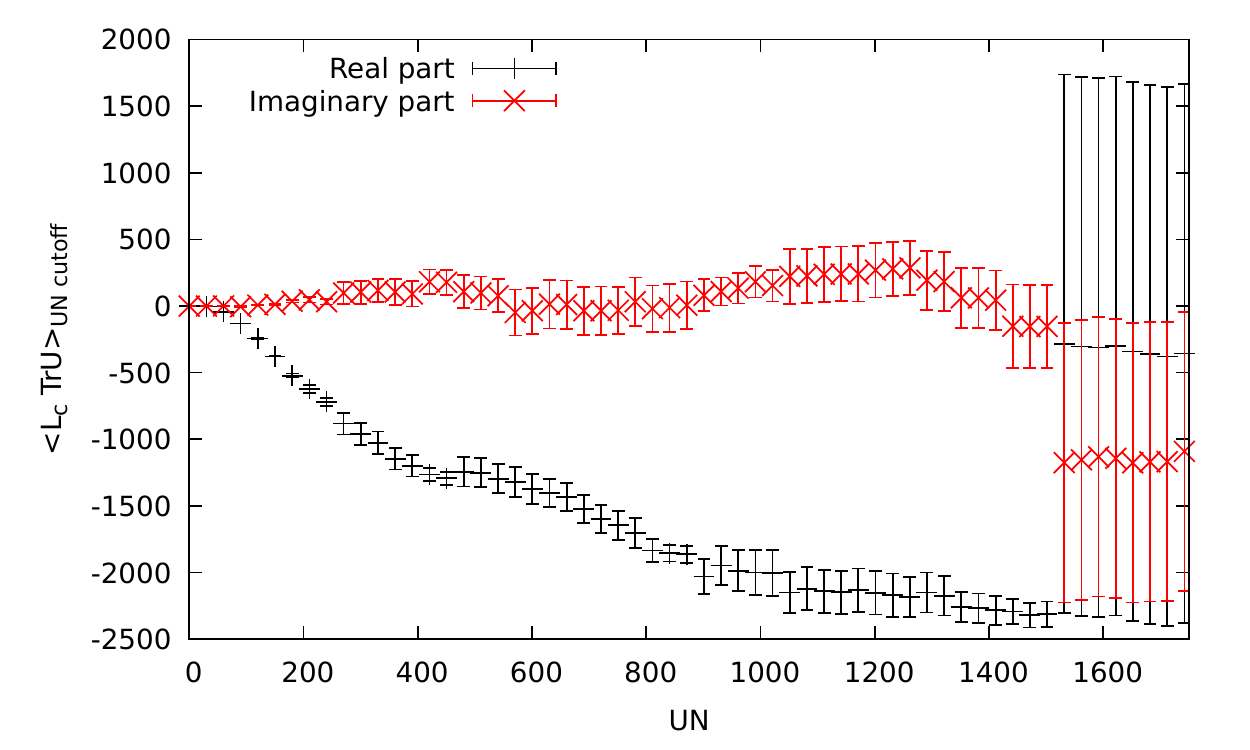} 
\caption{Boundary terms with (left) and without (right) gauge cooling for 
$\beta=2.0$, $\kappa=0.1$, $\mu=1.0$.} \label{fig:full} \end{figure} 
Fig.~\ref{fig:full} compares a simulation of the Polyakov chain with 
$N_t=16$ with and without gauge cooling. With gauge cooling the model is 
expected to give the correct results as the boundary terms go to zero. 
Thus it is no surprise that the exact value of 
$\left<\tr\cL\right>=2.0957$ is consistent with the CL simulation 
yielding $\left<\tr\cL\right>=2.0961(9)$. One can clearly see that without 
gauge cooling boundary terms develop. The simulation yields 
$\left<\tr\cL\right>=6.09(2)-0.04(1)i$, which is far off from the exact 
value.
 
\begin{figure}[h] \centering 
\includegraphics[width=0.45\textwidth]{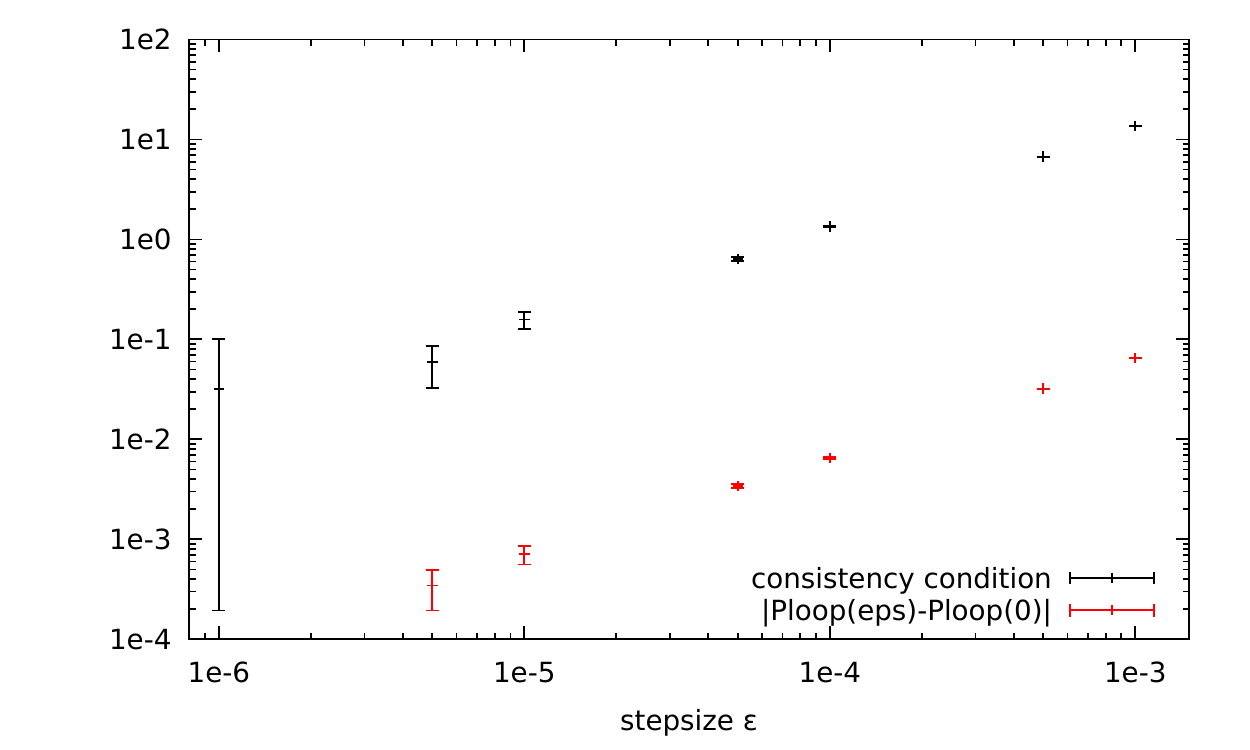} 
\caption{Step size dependence of $B(\infty)$ (consistency condition) and the average Polyakov loop with gauge cooling for $\beta=2.0$, 
$\kappa=0.1$, $\mu=1.0$.} \label{fig:steps} \end{figure}

For the full chain with parameters as above, there is a dependence on the 
step size $\epsilon$ in the value to which the boundary terms 
asymptotically seem to converge. $B(\infty)$ whose vanishing is the consistency condition (see equation \eqref{bound_cc}) fluctuates
strongly in 
the trajectories (between -100 and 100) hence even a tiny $\epsilon$ 
dependence effect is enhanced, see Fig \ref{fig:steps}, where we used 
fixed step size and the Euler-Maruyama discretization for the updates. The 
step size dependence goes with slope one in the double log plot, 
consistent with a linear dependence on $\epsilon$ as expected; note, 
however,  that the boundary term has a stepsize correction 
several orders of magnitude larger than the Polyakov loop itself.

\subsection{Heavy dense QCD (HDQCD)}
\label {ressec:hdqcd}
HDQCD was introduced originally in 
\cite{bender} for Wilson fermions and in \cite{blum} for staggered 
fermions. Later developments include \cite{feo,owe1,pr2014,owe2}. A first 
complex Langevin study was performed in \cite{Aarts:2008rr}; later gauge 
cooling as introduced in \cite{gaugecooling} and further developed in 
\cite{Aarts:2016qrv} has been used.

Complex Langevin for HDQCD produces a strong step size dependence when 
using the Euler-Maruyama discretization, just as in the Polyakov loop 
model of the previous section. Hence, to get away with larger step sizes, 
we use an improved updating method for the rest of this paper 
\cite{Ukawa:1985hr}. The boundary term for the Polyakov loop has the same 
form as in the Polyakov chain, see Eq.(\ref{pol.bound}).  

The boundary term for the plaquette looks similar; writing the 
plaquette as $\tr 
P\equiv \tr U_0 U_1 U_2^{-1} U_3^{-1}$, the boundary term integrand is: 
\begin{align} 
L_c\tr P&=-8\frac{N^2-1}{N}\tr P\nonumber\\
&+iK^{0}_a \tr\left(\lambda_a
U_0U_1U^{-1}_2U^{-1}_3\right)+iK^{1}_a \tr\left( 
U_0\lambda_aU_1U^{-1}_2U^{-1}_3\right)\nonumber\\
&-iK^{2}_a \tr\left(
U_0U_1U^{-1}_2\lambda_aU^{-1}_3\right)-iK^{3}_a \tr\left( 
U_0U_1U^{-1}_2U^{-1}_3\lambda_a\right)\,. 
\end{align}

\begin{figure}[h] 
\centering 
\includegraphics[width=0.45\textwidth]{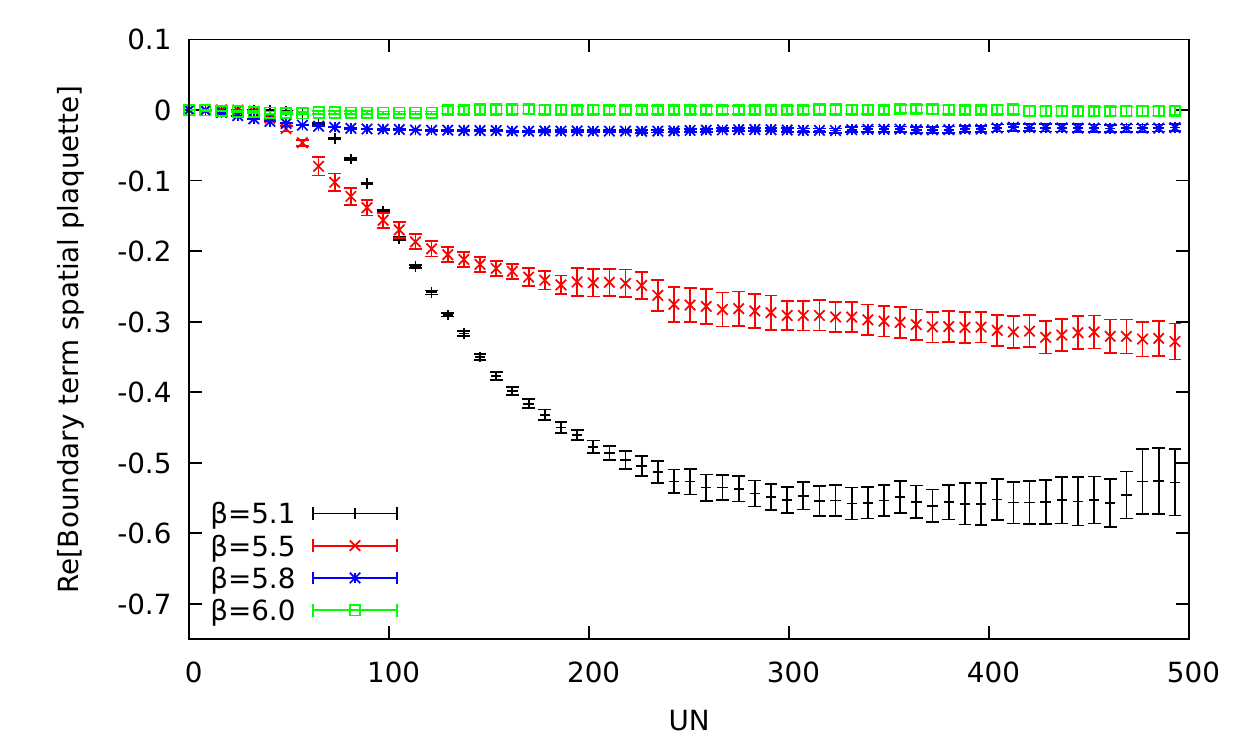} 
\includegraphics[width=0.45\textwidth]{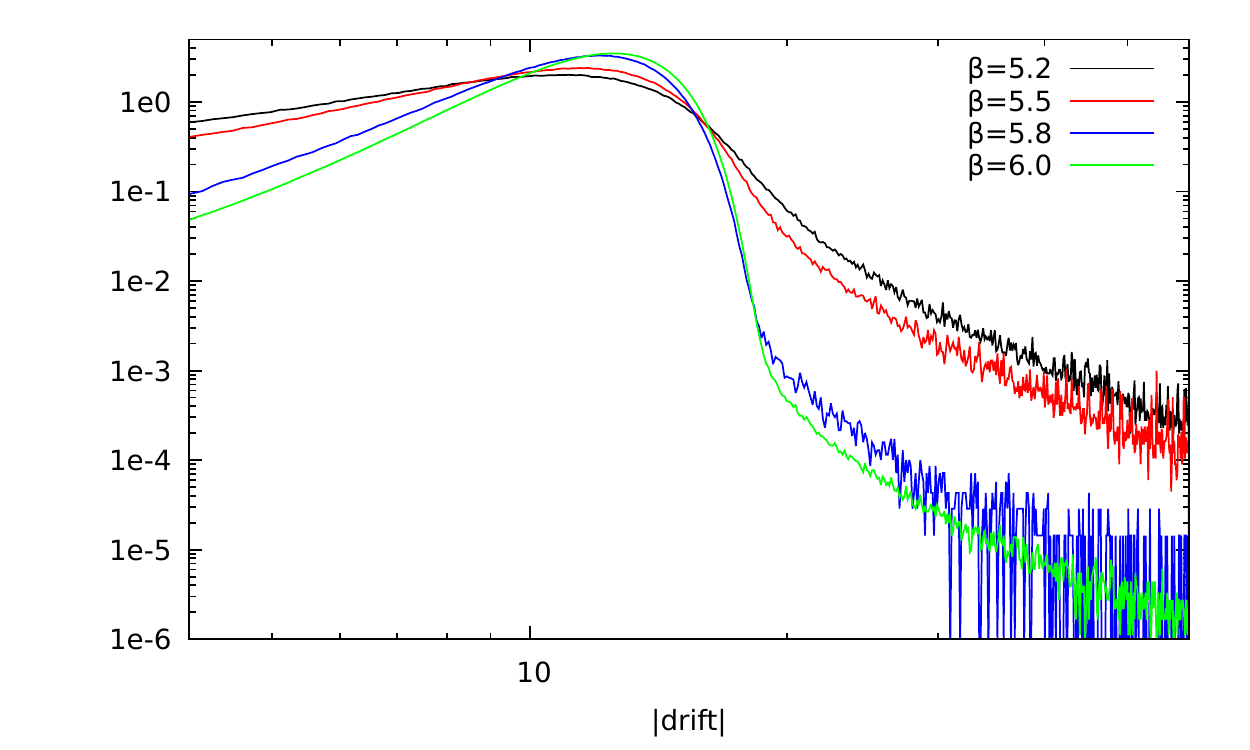} 
\caption{Left: Boundary term for spatial plaquettes in HDQCD. Right: 
Histogram for the absolute value of the drift terms in HDQCD. The histogram is plotted with a double-log scale. Parameters 
for both panels are: $6^4$ lattice with $\mu=0.85, \kappa=0.12,\ N_f=1$.} 
\label{fig:HDQCD} 
\end{figure}

Note that these formulas are the same for HDQCD and full QCD, the 
difference of the two theories are in the drift terms. For HDQCD correct 
results are accessible via reweighting, at least for not too large 
lattices.
Here we use for the cutoff the `unitarity 
norm' defined as
\bea 
UN= \max_{i,\mu} \tr (U_{i,\mu} U_{i,\mu}^\dagger -1)^2. 
\eea 
The results for the spatial plaquette average are shown in
the left panel of Fig.~\ref{fig:HDQCD}; we only 
show the plateau region of the boundary terms and leave out the region of
very large
unitarity norms because of large error bars. Boundary terms are present
even at $\beta=6.0$, though they become quite small in magnitude as $\beta$ 
increases. Note that in an earlier publication \cite{gaugecooling} it
was observed that the CL results are correct within errors
above $\beta \ge 5.8$. Here we collect averages in the long time
stationary phase of the system where a small deviation
develops also above $\beta \ge 5.8$. For these $\beta$ values at moderate
Langevin times one can see essentially correct results before
the rise of the unitarity norm signals the buildup of the boundary terms
measured here. This issue will be discussed in detail in an upcoming
publication \cite{transitionline}. 

In the right plot of figure \ref{fig:HDQCD} we show the criterion from 
\cite{Nagata:2016vkn}, which also shows that for   all $\beta$ CL is 
unreliable though for larger $\beta$ the tail in the distribution
  shrinks considerably. Thus both criteria are consistent. The boundary terms
  are directly related to the proof of convergence and lead to a 
  quantitative estimation of the magnitude of the error, see in Sec.~\ref{hdqcderror}. 

\begin{figure}[h] \centering 
\includegraphics[width=0.45\textwidth]{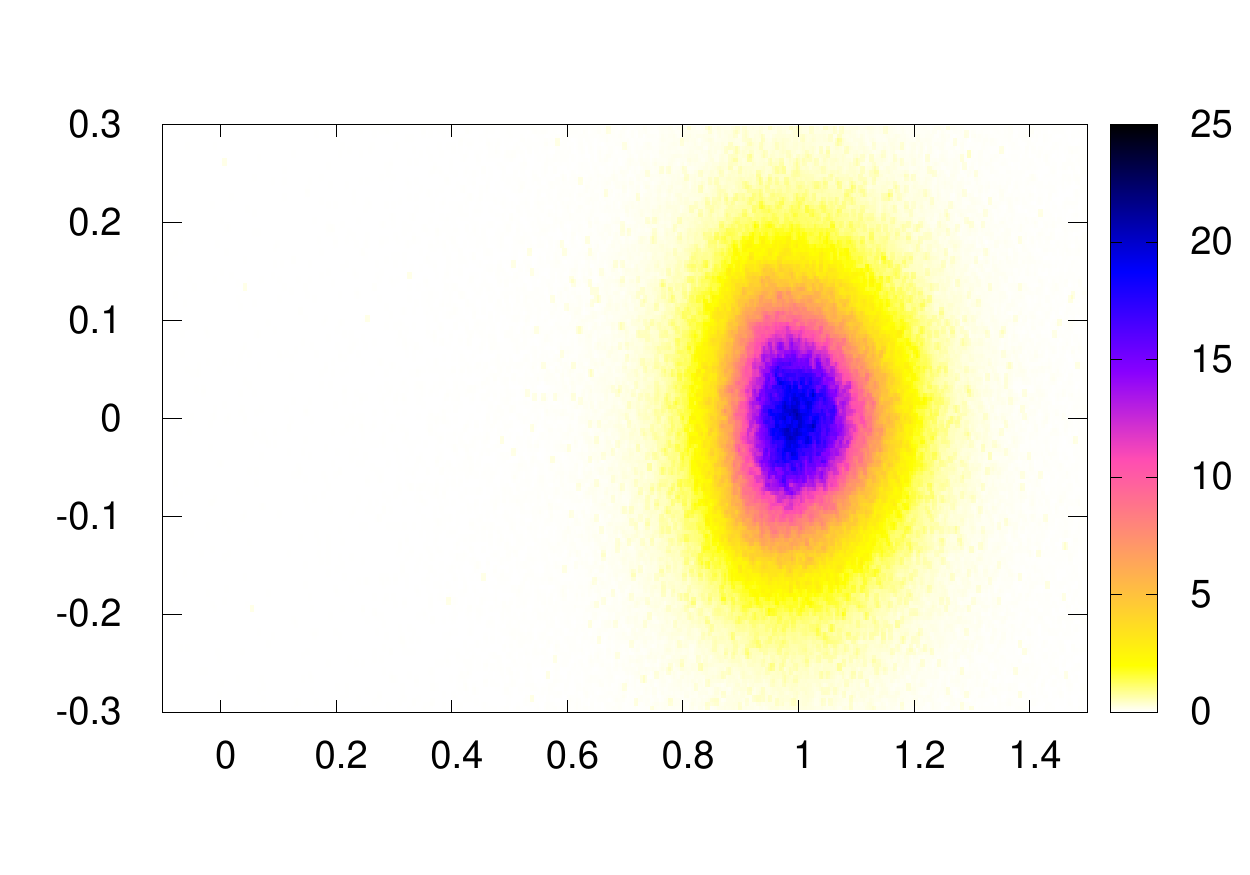} 
\includegraphics[width=0.45\textwidth]{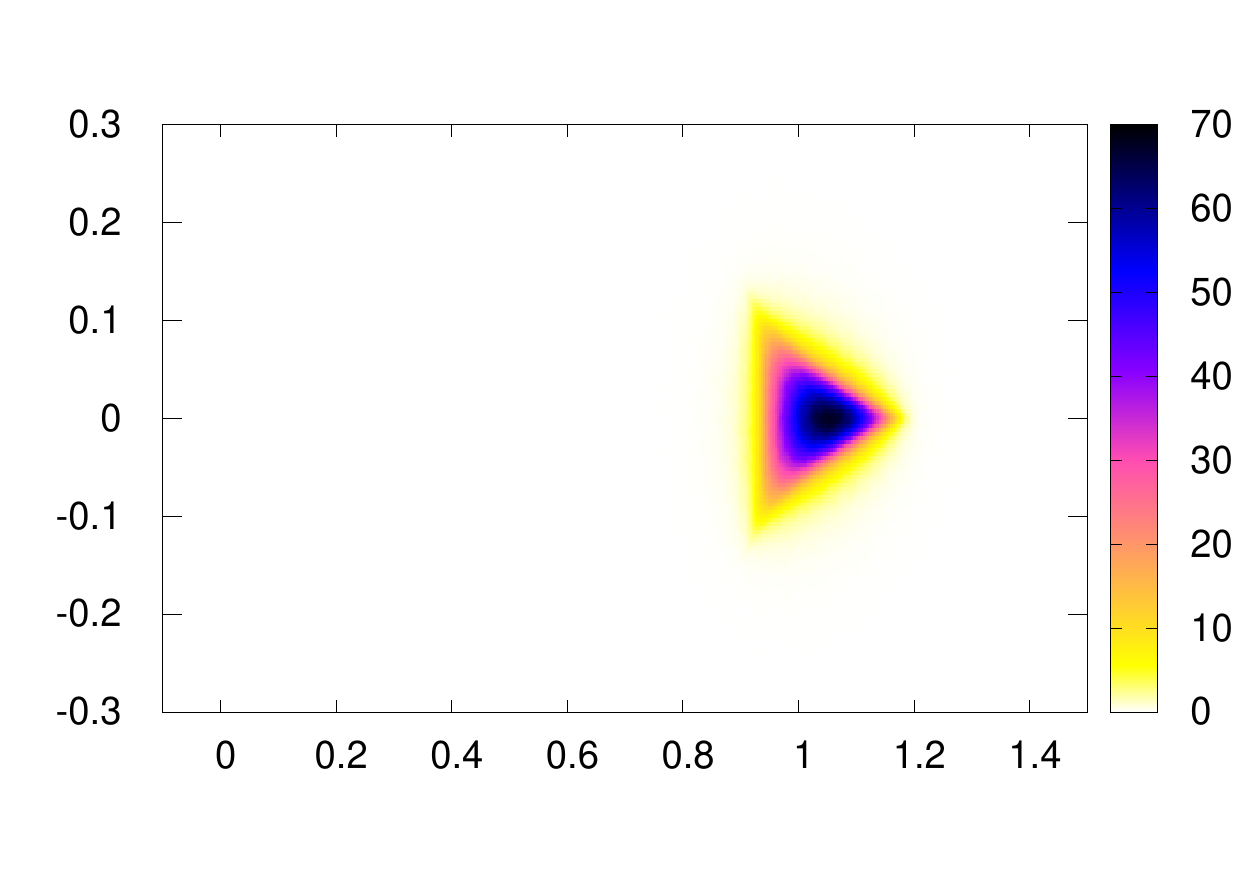} 
\caption{Histogram of the local determinants in HDQCD for $\beta=5.5$ 
(left) and $\beta=6.0$ (right). Other parameters as in 
Fig.~\ref{fig:HDQCD}} \label{fig:HDQCDdet} \end{figure}

Note that in HDQCD the determinant is a product over spatial sites of 
`local determinants'. In Fig.~\ref{fig:HDQCDdet} we show the histograms of 
the local determinants in the measure of HDQCD for $\beta=5.5$ and for 
$\beta=6.0$ in the CL simulation. One observes that the distributions are 
far away from zero, therefore at these parameter sets the zeroes of the 
measure on the complex manifold should have no measurable effect (such an effect is expected 
close to the critical chemical potential $\mu_{cr}=-\textrm{ln}(2\kappa)$ 
\cite{Aarts:2017vrv}).

The boundary terms for the Polyakov loop
appear, however, consistent with zero inside (albeit large) statistical errors,
even at the lower $\beta$ values where the average differs sizeably
from the reweighting result. We shall discuss this aspect further in Sec.~\ref{hdqcderror}.

\subsection{The 3D XY model} \label{sec:XY} 

Finally, we revisit the 3D XY model in which complex Langevin famously 
fails already for small imaginary parts of the action \cite{Aarts:2010aq}. 
The CL application to this model was analyzed carefully
in \cite{Aarts:2010aq}. It is of 
particular interest here because it shows that the occurrence of boundary 
terms depends on the observable considered.

The action reads 
\begin{equation} 
S=-\beta\sum_x\sum_{\nu=0}^2\text{cos}\left(\phi_x-\phi_{x+\hat{\nu}}-  
i\mu\delta_{\nu,0}\right)\,. 
\end{equation} 

\begin{figure} 
\includegraphics[width=0.45\textwidth]{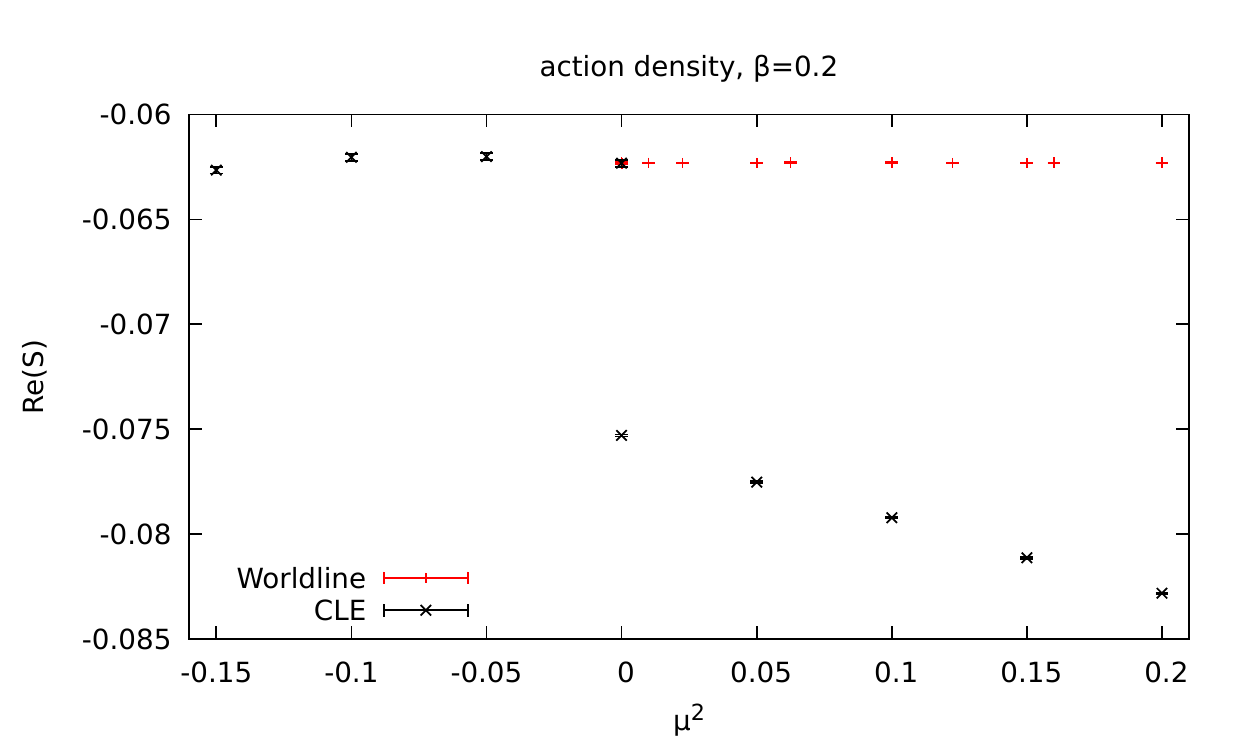} 
\includegraphics[width=0.45\textwidth]{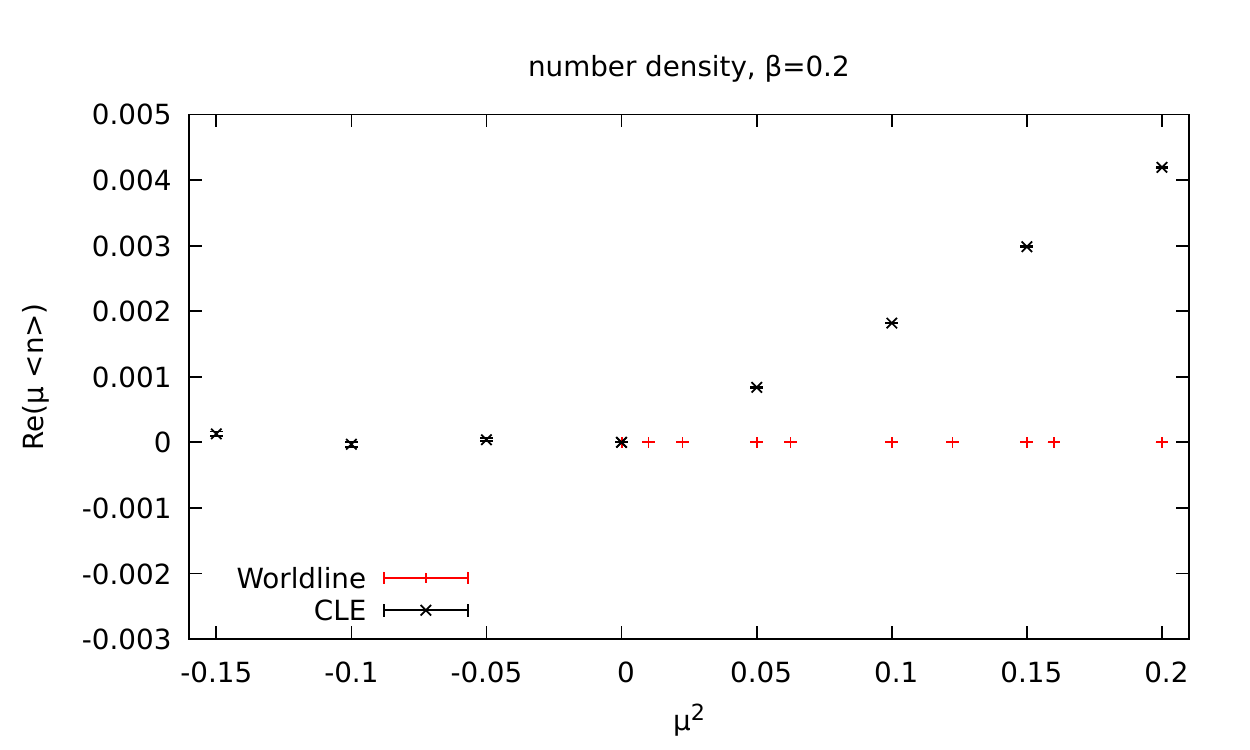} 
\includegraphics[width=0.45\textwidth]{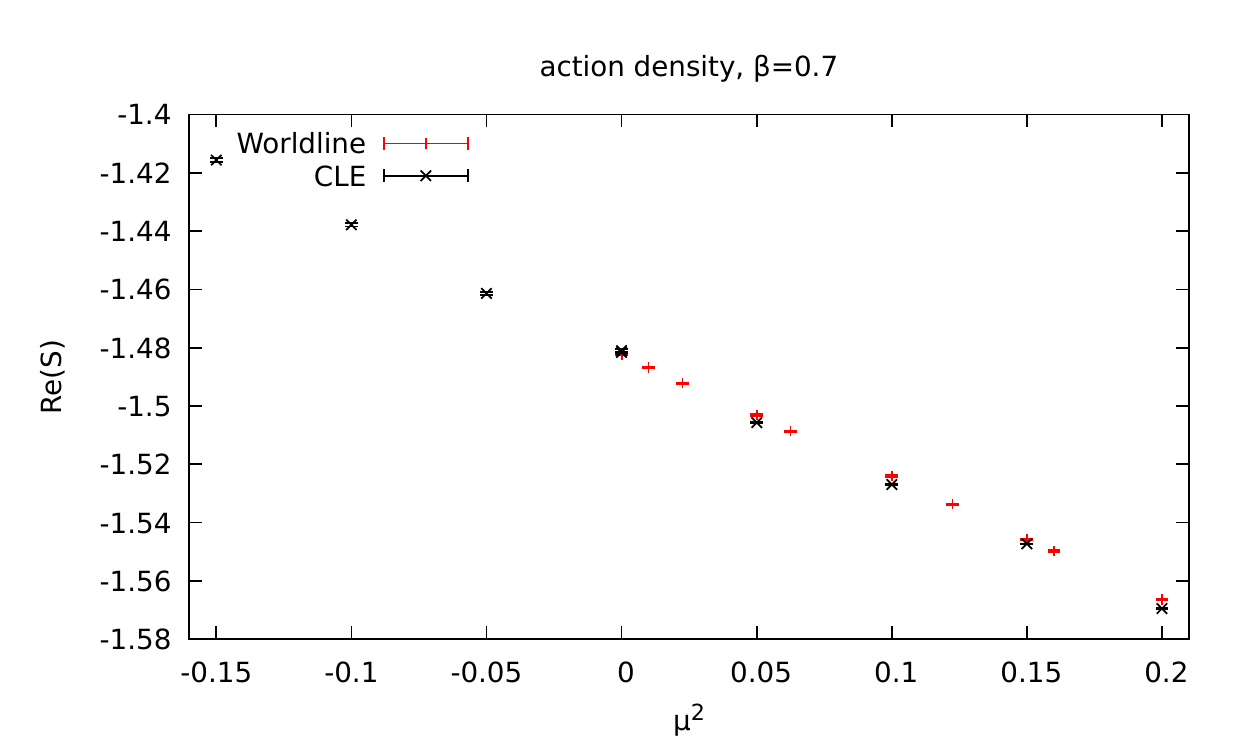} 
\includegraphics[width=0.45\textwidth]{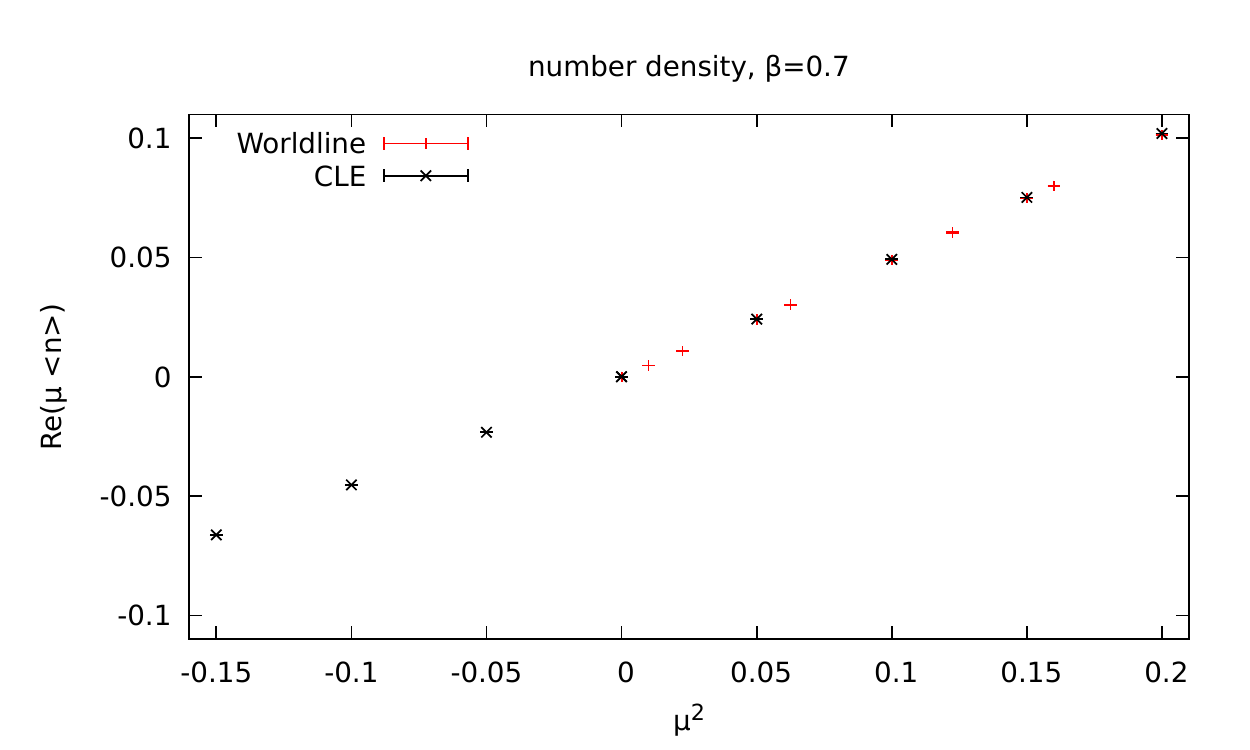} 
\caption{Action density (left) and number density (right) for $\beta=0.2$ 
(top) and $\beta=0.7$ (bottom). Note the apparent discontinuity at 
$\mu=0$ from the CL simulation in the action density, which does not 
show up for the number density.} 
\label{fig:XY} 
\end{figure}

Since there are no poles in this model the 
wrong convergence in complex Langevin can only come from boundary 
terms at infinity. We investigate two observables, the action density
\begin{equation} 
\langle S \rangle =-\beta\frac{\partial \text{ln} Z}{\partial \beta}\,, 
\end{equation} 
and the number density 
\begin{equation} 
n=\frac{\partial \text{ln} 
Z}{\partial \mu}= \left\langle i\beta\sum_x\text{sin}(\phi_x-\phi_{x+\hat{0}}-i\mu)\, \right\rangle. 
\end{equation} 
In the case of the action density as a function of $\mu^2$ it has been 
shown that for small $\beta$, CL produces a discontinuity at $\mu^2=0$ 
\cite{Aarts:2010aq}. We also show this in Fig.~\ref{fig:XY}, where we 
compare CL simulations with a worldline formulation\cite{Banerjee:2010kc},
which leads to correct results and thus is used as a benchmark for CL.

The discontinuity of the CL results for the action density can be 
understood as follows (see \cite{Aarts:2010aq}): at imaginary $\mu$, 
including $\mu=0$, when using real fields initially (`cold start'), the 
imaginary part of all drift terms are zero (even in the presence
of rounding errors on the computer) and thus the configuration 
remains real at all Langevin times. The process is thus equivalent to 
a real Langevin process, producing correct results and no boundary 
terms. For real nonzero $\mu$, no matter how small, the process will always
wander into the complexified configuration space, converging to an
equilibrium distribution extending into the complexification and boundary
terms can appear.

For $\mu=0$, however, there is a subtlety: a cold start will produce a 
real Langevin process and yield a result smoothly connected to those 
for purely imaginary $\mu$, as stated above; on the other 
hand, starting with an initial configuration with nonzero imaginary parts 
(`hot start'), the process will explore the complexified configuration 
space, converge to an equilibrium distribution not supported on the real 
subspace, producing a result smoothly connected to those for real $\mu\neq 
0$ and boundary terms can appear. To make sure that we test the boundary terms in the complexified distribution we always use a positive chemical potential below.

Note however that in the number density there is no apparent 
discontinuity at $\mu=0$ (which reflects the fact that the real part of the density is proportional to $\textrm{sinh}(\mu)$), unlike in the action density, see Fig.~\ref{fig:XY}. 
The discontinuity in the action density disappears for larger $\beta$ and 
complex Langevin apparently leads to correct results. We will investigate 
this further by means of boundary terms below.\\ 
In the formulas below we will use the shorthands
\bea
\nabla_x  = \nabla_{\phi_x}, \quad  K(x) = K(\{\phi_x\}).
\eea
For the boundary term of the action density we need

\begin{equation} L_c 
S =\left(\nabla_x+K(x)\right) \nabla_x S=-\nabla_x K(x) -K^2(x)\,, 
\end{equation} 
with 
\begin{align} K(x)=&-\nabla_x S= \beta\sum_{\nu=0}^2 
\left[-\text{sin}(\phi_x-\phi_{x+\hat{\nu}}- 
i\mu\delta_{\nu,0})+\text{sin}(\phi_{x-\hat{\nu}}-\phi_{x}- 
i\mu\delta_{\nu,0})\right]\\
\nabla_x K(x)=&\beta\sum_{x,\nu=0}^2\left[-\text{cos}(\phi_x- 
\phi_{x+\hat{\nu}}-i\mu\delta_{\nu,0})-\text{cos}(\phi_{x-\hat{\nu}}- 
\phi_{x}-i\mu\delta_{\nu,0})\right]\,. 
\end{align} 
For the number density we need 
\begin{equation} 
L_c n =\left(\nabla_x+K(x)\right) \nabla_x n=\nabla^2_x n +K(x)\nabla_x n \,, 
\end{equation} 
with $K(x)$ 
as before and 
\begin{align} 
\nabla_x n &=i\beta\left( \text{cos} 
(\phi_x-\phi_{x+\hat{0}}-i\mu)- 
\text{cos}(\phi_{x-\hat{0}}-\phi_{x}-i\mu)\right)\\
\nabla^2_x 
n &=i\beta\sum_x\left( -\text{sin} (\phi_x-\phi_{x+\hat{0}}-i\mu)-
\text{sin}(\phi_{x-\hat{0}}-\phi_{x}-i\mu)\right)\,. 
\end{align} 

\begin{figure}[ht] 
\includegraphics[width=0.45\textwidth]{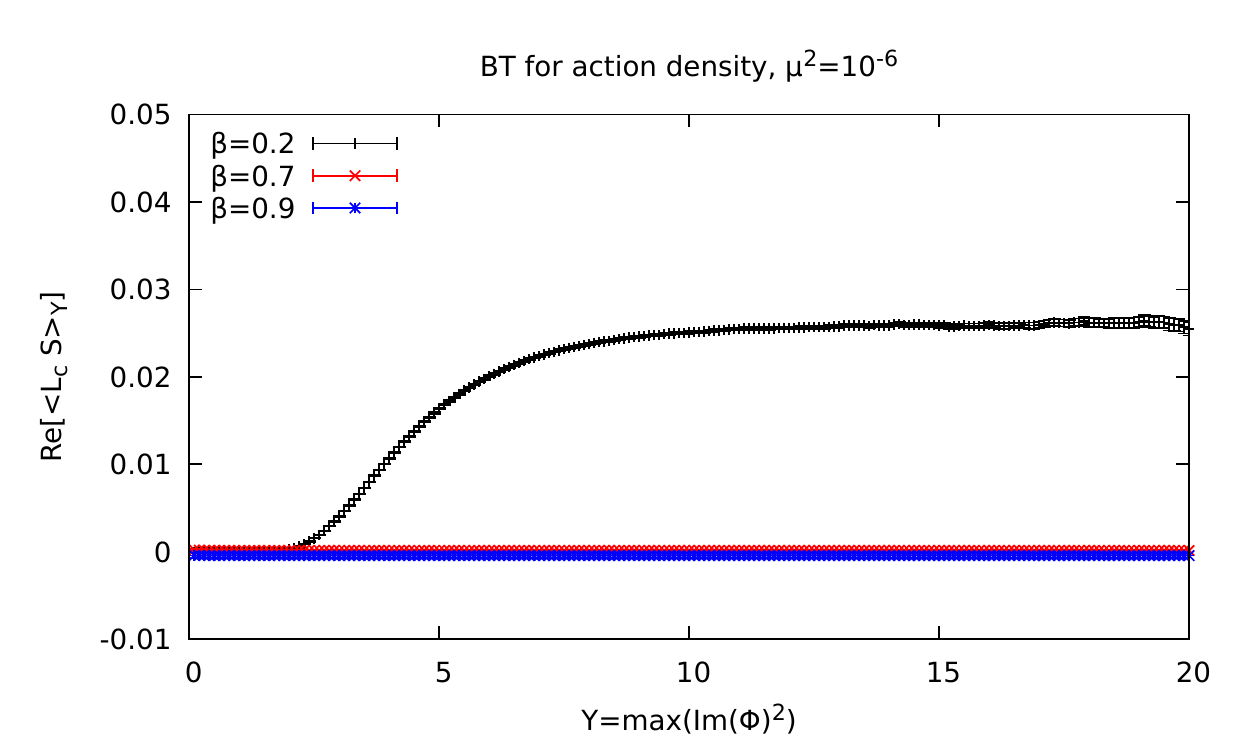} 
\includegraphics[width=0.45\textwidth]{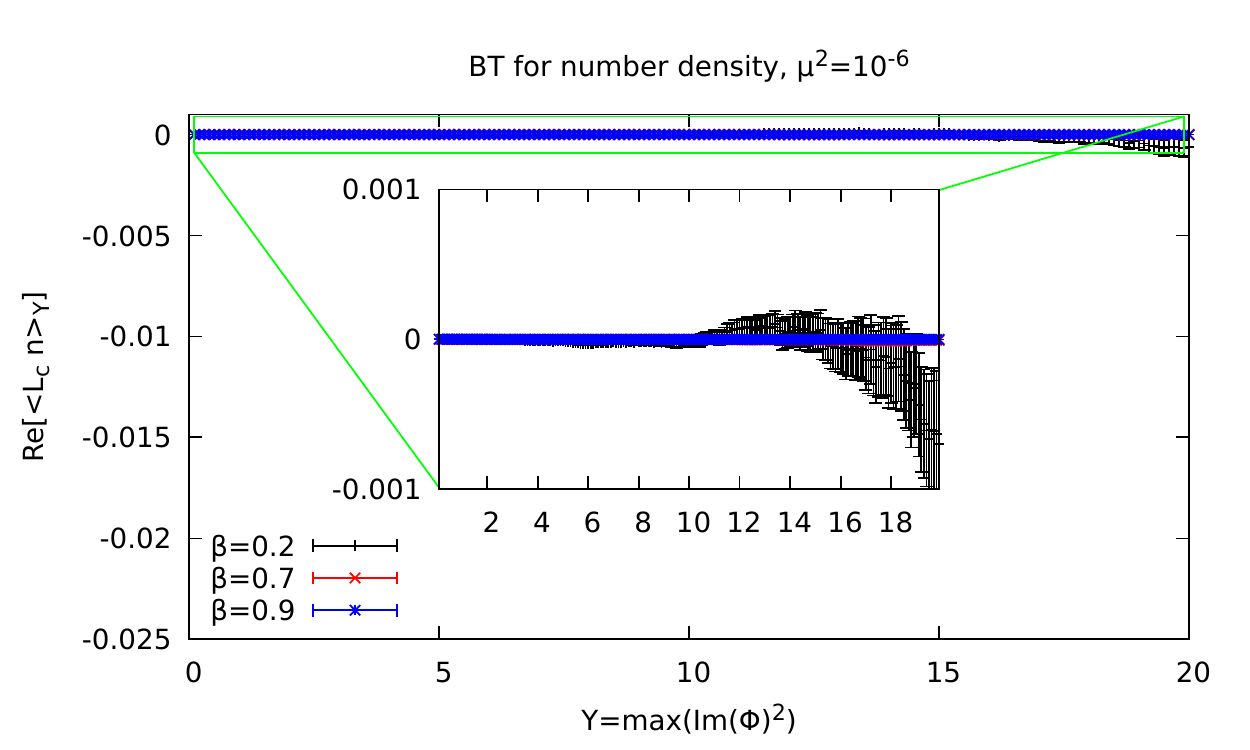} 
\includegraphics[width=0.45\textwidth]{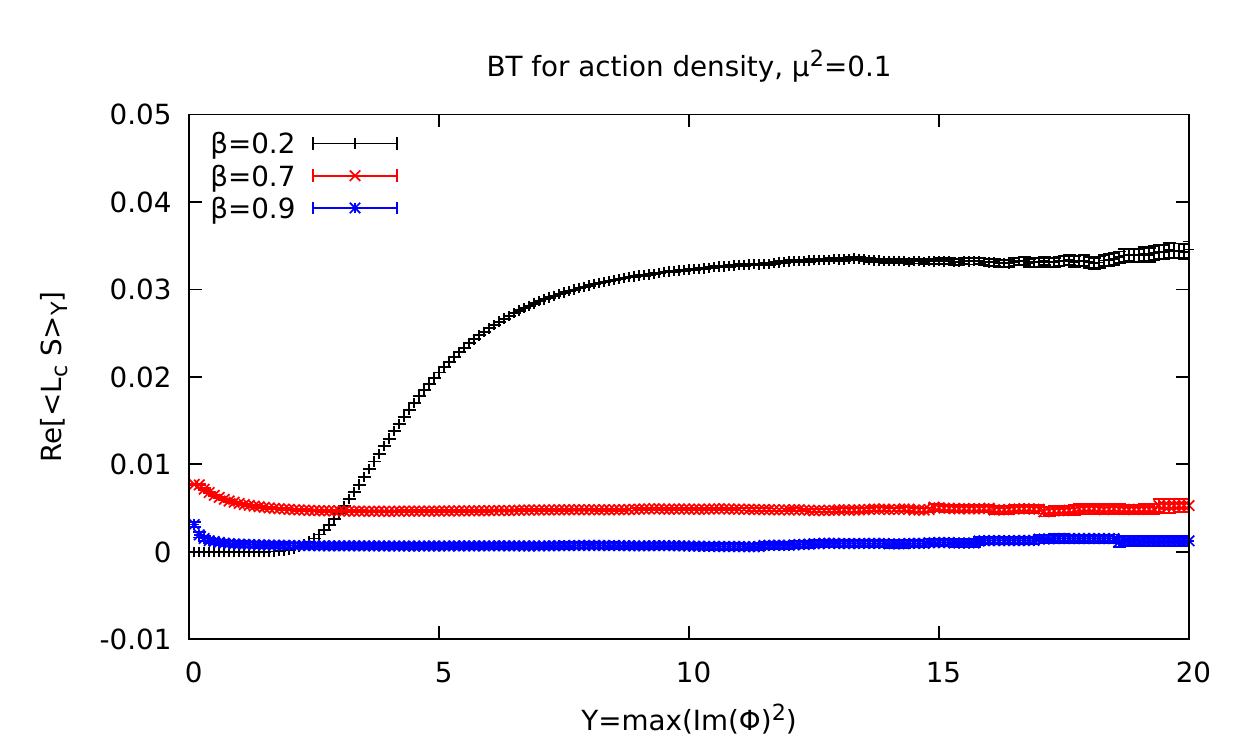} 
\includegraphics[width=0.45\textwidth]{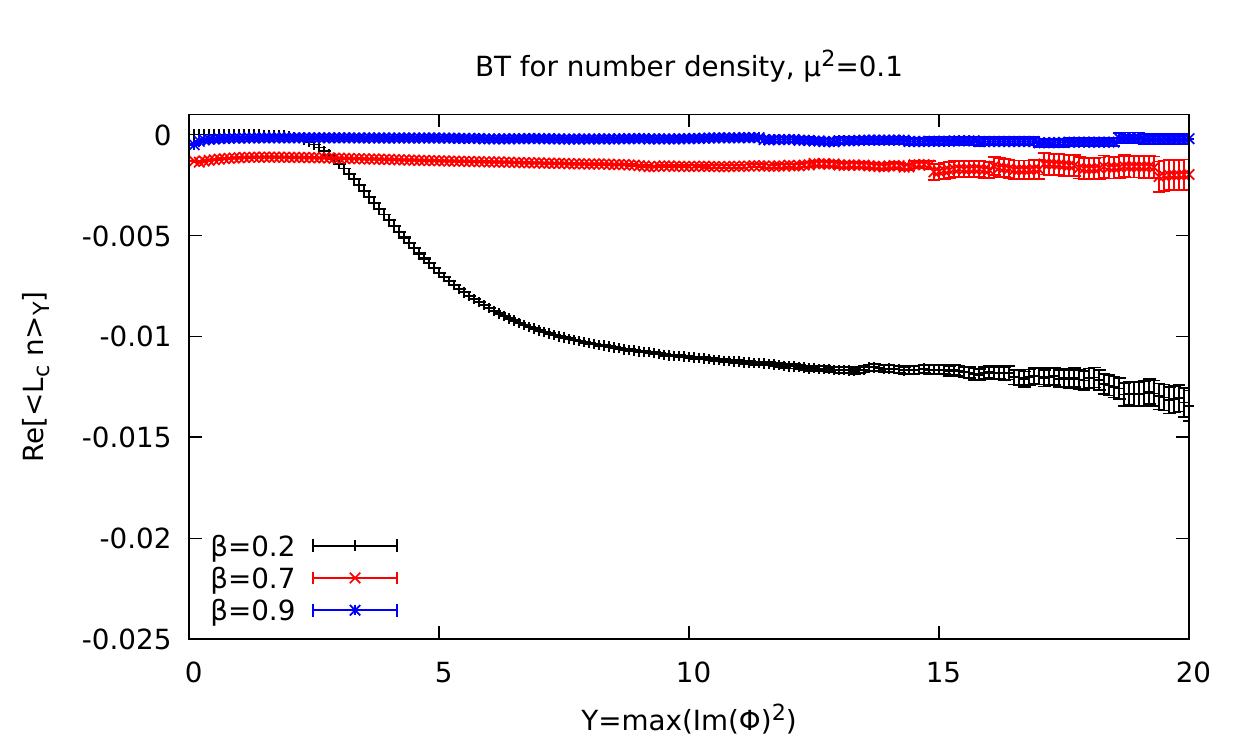} 
\includegraphics[width=0.45\textwidth]{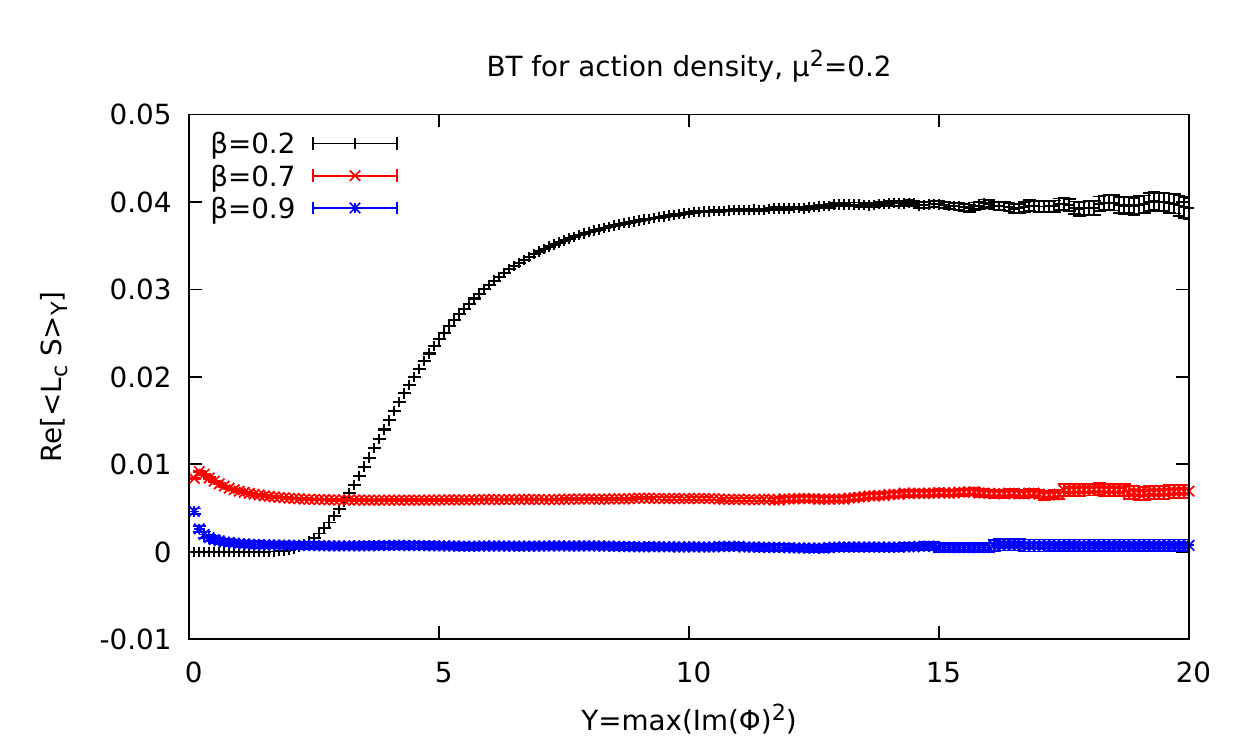} 
\includegraphics[width=0.45\textwidth]{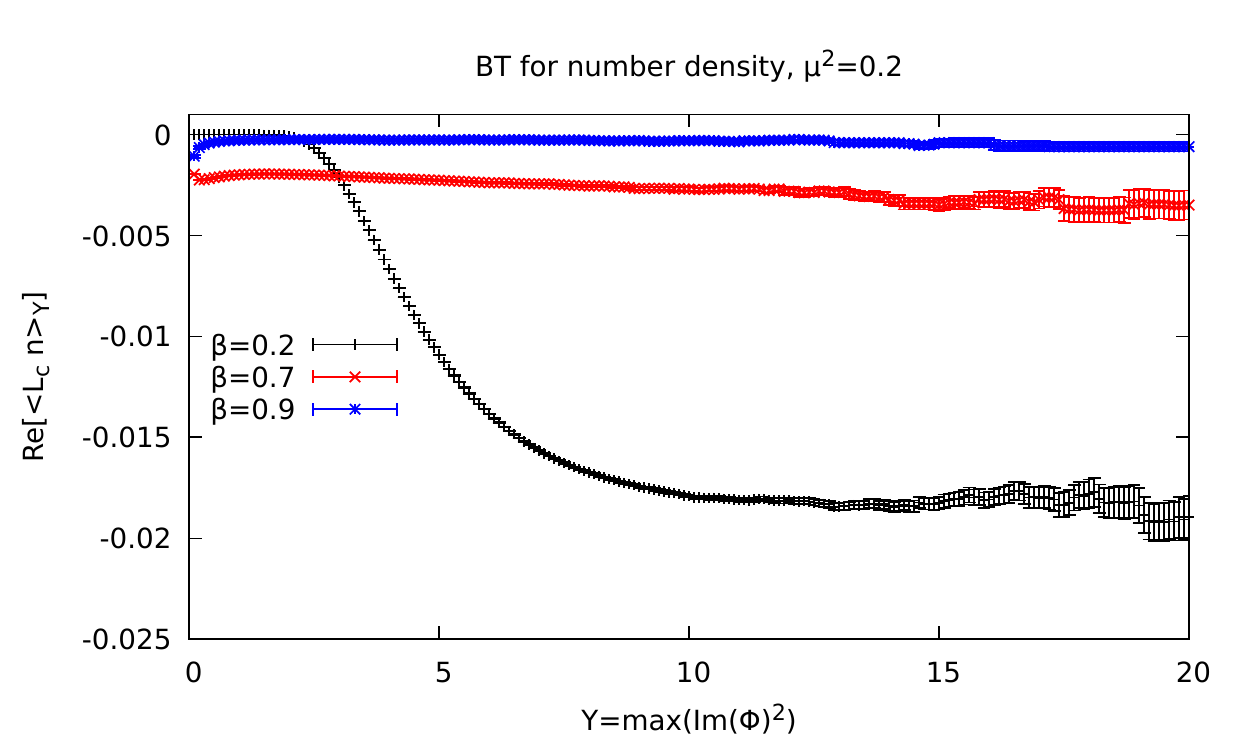} 
\caption{Boundary terms (called $B_1$ in section \ref{sec:est}) in the XY model. Left column: BT for action 
density. Right column: BT for number density. Top to bottom: 
$\mu^2=10^{-6},0.1,0.2$. } 
\label{fig:XY_BTS} 
\end{figure}

We computed the boundary terms for both observables for $\beta=0.2,0.7,0.9$ 
and $\mu^2=10^{-6}, 0.1,0.2$; the results are shown in Fig.~\ref{fig:XY_BTS}. 

In the case of $\mu^2=0.1,0.2$ we find that the boundary terms for both 
observables are largest for $\beta=0.2$ as expected.

The action density has non-vanishing boundary terms for all three values of 
$\mu$ and $\beta=0.2$ and $0.7$. There is still a tiny, barely visible 
boundary term even for $\beta=0.9$. Hence, we conclude that in this model 
the action density always has some boundary terms which can become 
arbitrarily small as $\beta$ increases. The observables for $\beta=0.9$, 
however, can be regarded as correct for all practical purposes.
The number density, on the other hand, has no boundary terms at 
$\mu^2=10^{-6}$, for all the three $\beta$ values studied.
This demonstrates that the inclusion of the 
observable is crucial in the computation of boundary terms. The 
correctness of the CL evolution does not only depend on the distribution 
of the drift.
The vanishing of the boundary terms for the number density is consistent  
with the lack of an apparent jump in Fig.~\ref{fig:XY}.

Note that in Fig.~\ref{fig:XY_BTS} we again only show the boundary term
up to the end  of the plateau-like 
region. For larger values of $Y$ huge error bars start to appear due to 
statistical outliers which typically lead to large values in the boundary 
term. Those outliers also sometimes lead to sudden jumps and larger 
errorbars, see e.g. the red curve in the center right plot of Fig.~\ref{fig:XY_BTS}. The identification of a plateau-like region is enough to 
identify a boundary term since the limit $Y\rightarrow\infty$ can be taken 
by extrapolation. Values beyond the plateau like region, where the error 
bars become very large should be discarded. The last point, which includes 
all $Y$ should always be consistent with zero, since this is nothing but 
the `consistency condition' from \cite{Aarts:2011ax},
signifying that the process has equilibrated. We checked that this 
is the case in all our simulations.

Finally we also look at the drift criterion from \cite{Nagata:2016vkn}. 
\begin{figure} 
\includegraphics[width=0.45\textwidth]{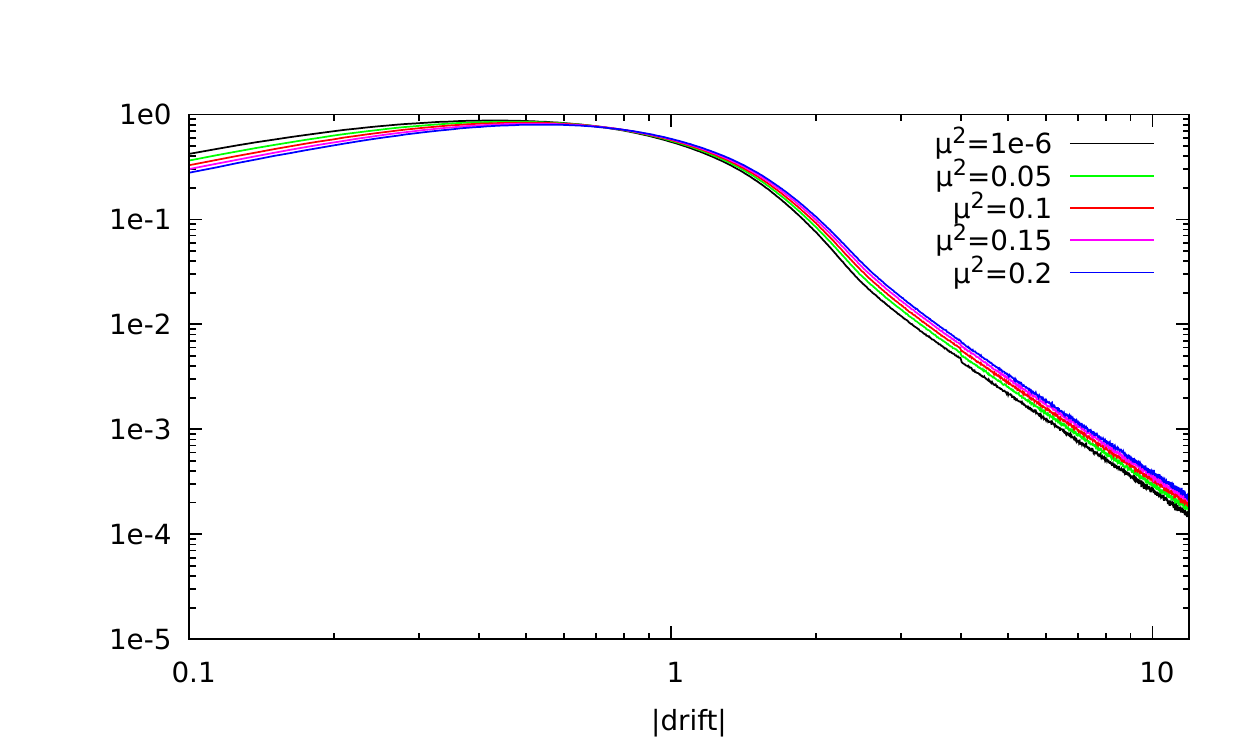} 
\includegraphics[width=0.45\textwidth]{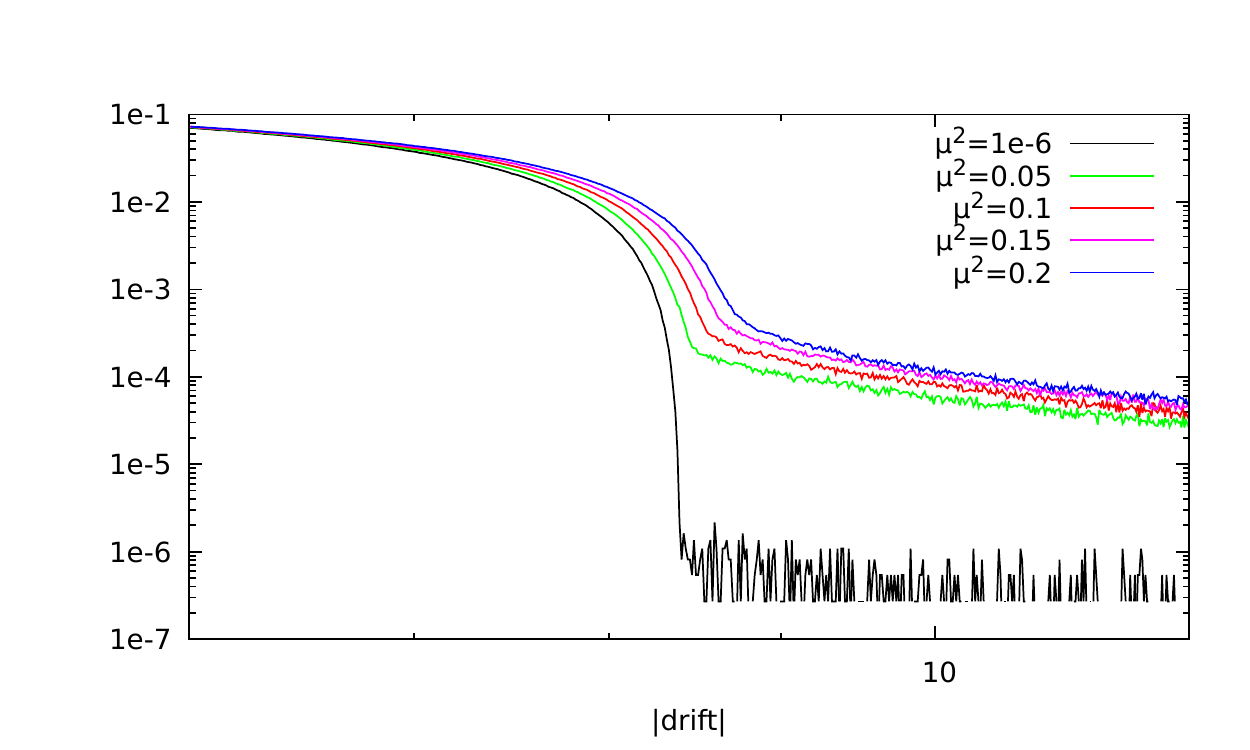} 
\includegraphics[width=0.45\textwidth]{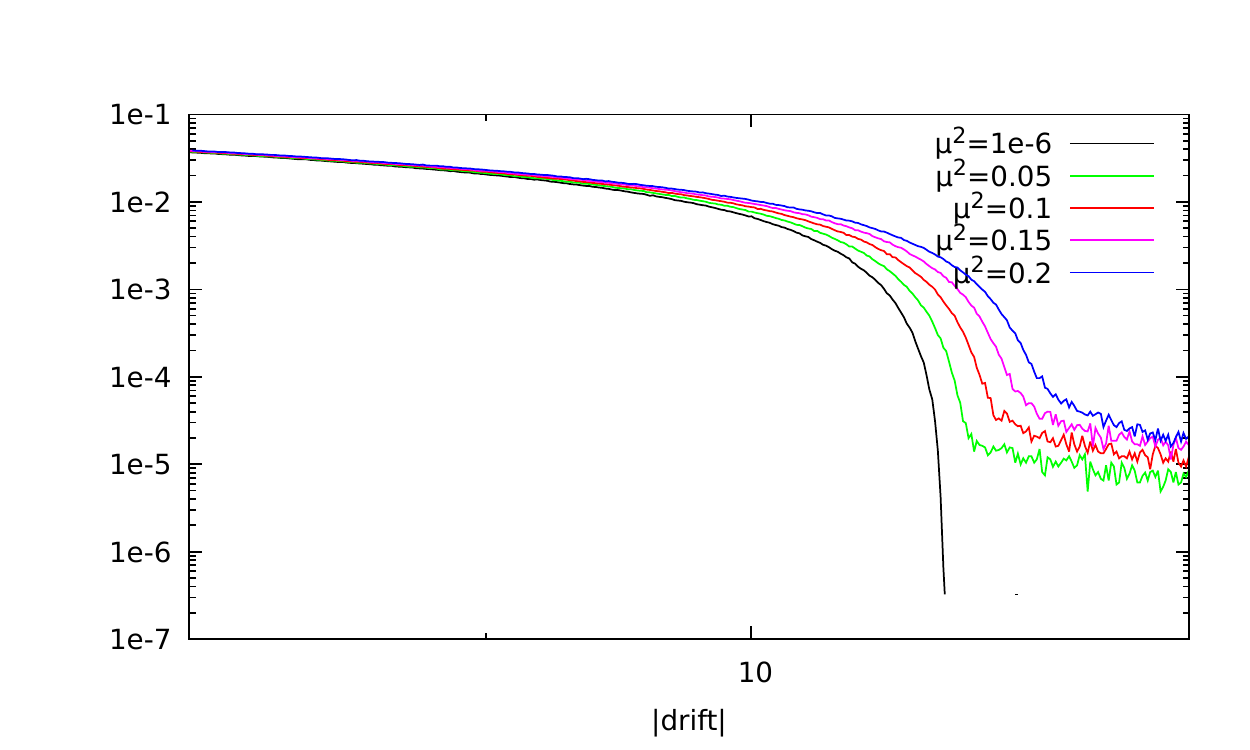} 
\caption{Histogram of the drift for $\beta=0.2$ (top left), $\beta=0.7$ 
(top right) and $\beta=0.9$ (bottom) in the XY model. All plots have a double-log scale.} 
\label{fig:drift_hist_XY} 
\end{figure} 
Fig.~\ref{fig:drift_hist_XY} shows the histogram of the absolute value 
of the drift. The drift criterion predicts that for $\beta=0.2$ results 
are wrong, the same is true for $\beta=0.7,0.9$ 
for $\mu^2>10^{-6}$. For $\mu^2=10^{-6}$ and $\beta=0.7$ the tail is strongly suppressed, suggesting a small error of CL, while for $\beta=0.9$ there is no tail at all, suggesting that CL is correct here.
While this criterion does show the same 
sensitivity and also signals slightly wrong convergence for 
$\beta=0.7,0.9$, it does not take into account the observable and thus 
cannot find that the number density for $\mu^2=10^{-6}$ is actually 
correct for all $\beta$, while the action density is not.
 Since this criterion relies on the interpretation of the behavior
  of a distribution in the region of small values and large errors, it is
  more of a qualitative nature and would not allow a quantitative estimate of
  the deviation of the CL results from the exact ones.

\section{Estimation of the systematic error of CL from boundary term analysis}
\label{sec:est}

The systematic error of the CL result is given by
\bea \label{clerror}
F_{\cO}(t,0)-F_{\cO}(t,t) = \bra \cO \ket_{P(t)}- \bra \cO \ket_{\rho(t)}\,.
\eea
Calculating this difference would allow us to get the exact result, however
generally $F(t,\tau)$ is not directly accessible for $\tau>0$, except for simple
toy models.

The time evolved observable is
\bea
\mathcal{O}(z;\tau) = \exp (\tau L_c) O(z).
\eea
Assuming that the spectrum of $L_c$ is discrete and contained in the open left half plane -- except for a simple eigenvalue at zero -- we have
\bea
\mathcal{O}(z;\tau) = \sum_{n=0}^\infty a_n(z) \exp (-\omega_n \tau ) 
\eea
where $a_0$ is independent of $z$ and 
\bea
 \omega_0=0; \quad \textrm{Re}\omega_n >0 \textrm{ for } n>0.
 \eea
 We are interested in $a_0$ which gives the correct expectation value:
 \bea
 a_0= \lim_{t \rightarrow \infty} F_\mathcal{O}(t,t)= \lim_{\tau \rightarrow \infty}
 \int dx dy P(x,y;0) \mathcal{O}(x+iy;\tau)
 \eea
 In general we have
 \bea \label{fttexact}
 F_\mathcal{O}(t,\tau) =  \sum_{n=0}^{\infty} A_n(t) \exp( -\omega_n \tau)
 \eea
 with
 \bea
 A_n(t) = \int dx dy P(x,y;t) a_n(x+iy).
 \eea
 To relate $F_\mathcal{O}(t,0)$ to $F_\mathcal{O}(t,t)$ we use a simplified
 ansatz based on the first two terms in (\ref{fttexact}):
\bea \label{fttansatz}
F_\mathcal{O}(t,\tau)= A_0+ A_1 e^{-\tau \omega_1  }.
\eea
  This ansatz in consistent with the assumption that the $\tau$ derivative of $F(t,\tau)$ is maximal at $\tau=0$.
 In \cite{boundaryterms1} we have calculated $F(t,\tau)$ for the $U(1)$ one
 plaquette model, and we have seen that this ansatz is a good description of the
 full $F_\mathcal{O}(t,\tau)$.
 This leads to $F_\mathcal{O}(t,0)-F_{\mathcal{O}}(t,t)=A_1 $ for large $t$, where we denote by $A_1$ the limit of $A_1(t)$ for large $t$. We can access the
 constants $A_1, \omega_1 $ at $\tau=0$ by calculating
\bea
\left. { \partial^n F_\mathcal{O}(t,\tau) \over \partial \tau^n } \right|_{\tau=0} = B_n ,
\eea
where $B_1$ is what we called the boundary term above. Using
the ansatz one sees that $ B_1 = - \omega_1 A_1 ,\  B_2 = \omega_1 ^2 A_1 $, and finally
the systematic error of CL (\ref{clerror}) is given by $ A_1 = B_1^2 / B_2$.
One can show that in the CL process the boundary terms are calculated by 
\bea
B_n = \lim_{Y\rightarrow \infty} \int_{-Y}^{Y} P(x,y,t) L_c^n \mathcal{O}(x+i y) dx dy
\eea
with a reasoning similar to that leading to eq. (\ref{bound_cc}).
Having an estimate of the systematic
error allows us to calculate the corrected CL result:
\bea
\langle O \rangle_\textrm{corr} = \langle O \rangle_P - { B_1^2 \over B_2 }.
\eea
In the following Tables the column `CL error' (systematic error)
is calculated as:
$ \textrm{`CL error'}=\textrm{`CL'} - \textrm{`correct'}$ where `correct' comes
from other calculations considered as providing correct results such as
direct integration, reweighting (for a mild sign problem) and
the worldline setup. Thus ideally the
agreement between columns `CL error' and $ B_1^2/B_2 $ signals that
the ansatz (\ref{fttansatz}) describes the evolution
of $F(t,\tau)$ well and the corrected CL result will be accurate.

\subsection{U(1) one-plaquette-model}
\label{sec:estu1}

\begin{figure} 
\includegraphics[width=0.48\textwidth]{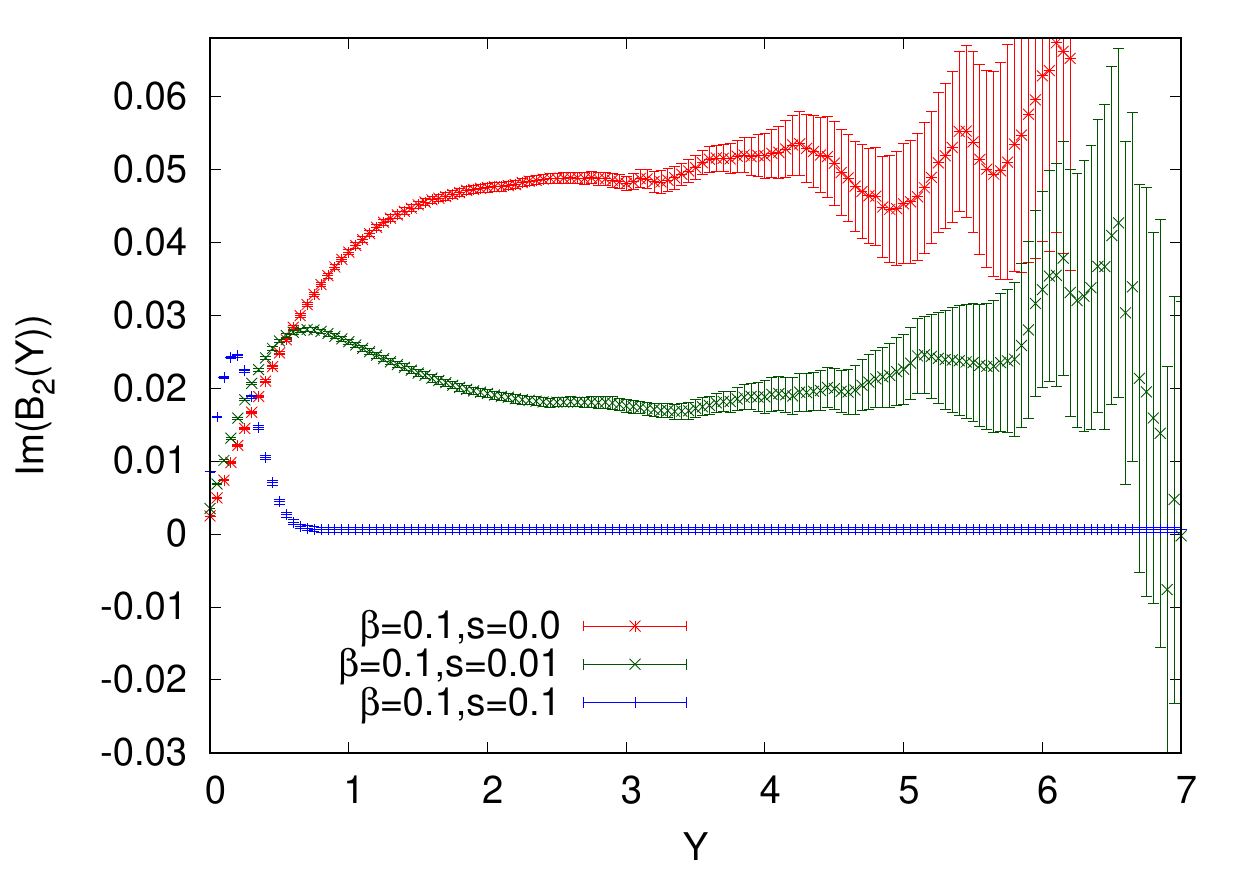} 
\caption{The imaginary part of the boundary term $B_2(Y)$ for the observable
  $e^{i x}$ is shown as a function of $Y$ for $\beta=0.1$ and
  several $s$ values in the $U(1)$ one-plaquette model.
 See also Fig.~\ref{fig:ycc1} for the boundary term $B_1(Y)$.
  } \label{fig:ycc1-b2}
\end{figure}

In Fig.~\ref{fig:ycc1-b2} we show the imaginary part of boundary term $B_2$ as a function of the cutoff $Y$. The corresponding formula is shown in App.~\ref{sec:eqappendix}.
Note the similarity
with Fig.~\ref{fig:ycc1}, except for the inverted sign and much larger fluctuations present in $B_2(Y)$. As shown in Appendix B of \cite{boundaryterms1},
it is expected that $\omega_1 \approx 1$ for this model, which amounts to $ B_1 \approx -B_2 $.
In the Table~\ref{u1table} we show estimations of the error of the CL method
for several parameter values.
\begin{table}
\begin{tabular}{|l|c|c|c|c|c|c|c|}
  \hline
  $\beta, s$ &  $B_1$ & $B_2$ & $B_1^2/B_2$ & CL error & CL & correct & corrected CL\\
  \hline
0.1,  0 & -0.04859(45)& 0.0493(11)& 0.04786(79) &  0.04891(45)  &  -0.00115(45)&-0.05006 &-0.04901(62)\\
0.1,  0.01 & -0.01795(49)& 0.01801(80)& 0.01789(60)& 0.01689(50) & -0.03318(50) & -0.05006 & -0.05106(40) \\
0.1,  0.1 & -0.00048(30)& 0.00057(35)& 0.00039(28)&  0.00049(31)  & -0.04957(31)     &  -0.05006 &-0.04997(6)\\
\hline
0.5,  0 & -0.2474(11) & 0.237(11)& 0.258(11)& 0.25818(23)& 0.00003(23)   & $-0.25815 $&  -0.258(11) \\
0.5,  0.3 & $-0.05309(86)$& 0.0552(51)&0.0507(41) & 0.04183(70)& -0.19658(70)   &$ -0.23841$ &  $-0.2473(37)$ \\
  \hline
\end{tabular}
\caption{The estimation of the systematic error of CL for the $U(1)$
  one plaquette model for the imaginary part of the observable $\mathcal{O}= e^{ix} $. }
  \label{u1table}
\end{table}
One notes that whitin errors, this method yields the correct value of the
systematic error due to boundary terms. Note that
using this estimate for the systematic error to correct the
CL results we get the exact result
within statistical errorbars.

\subsection{The 3D XY model}
\begin{figure}[ht] 
\includegraphics[width=0.45\textwidth]{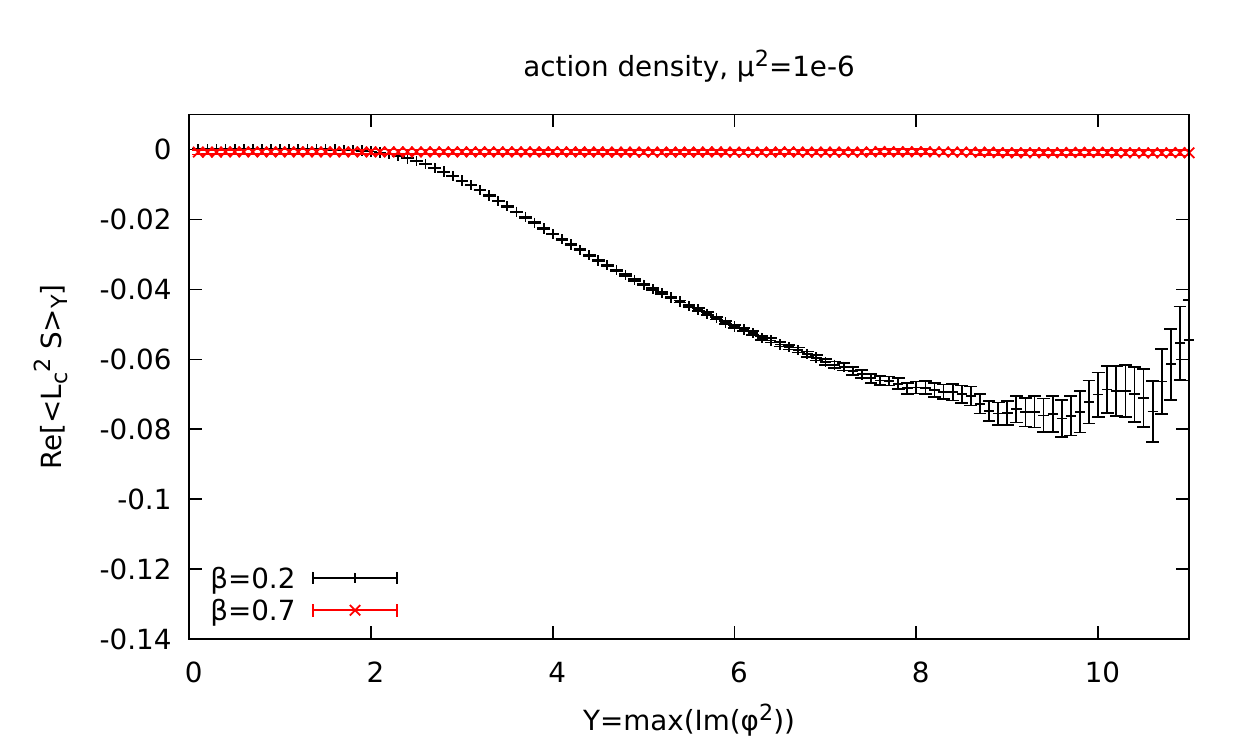} 
\includegraphics[width=0.45\textwidth]{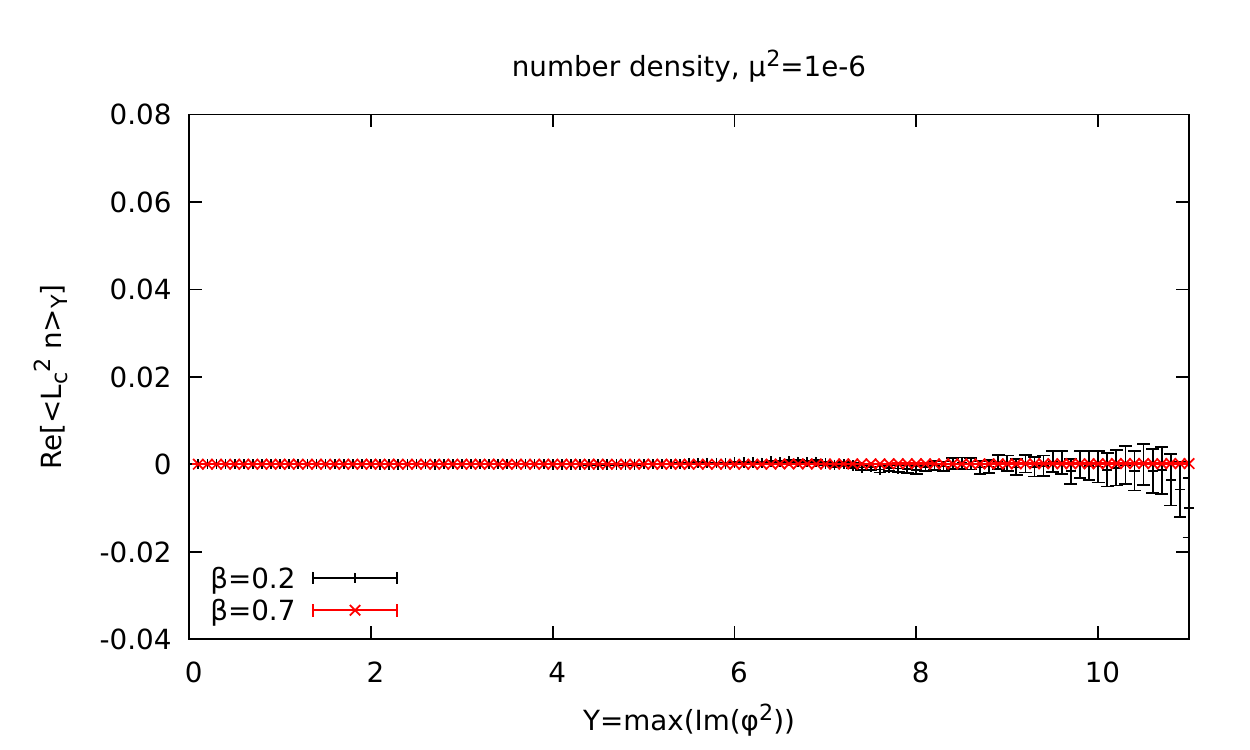} 
\includegraphics[width=0.45\textwidth]{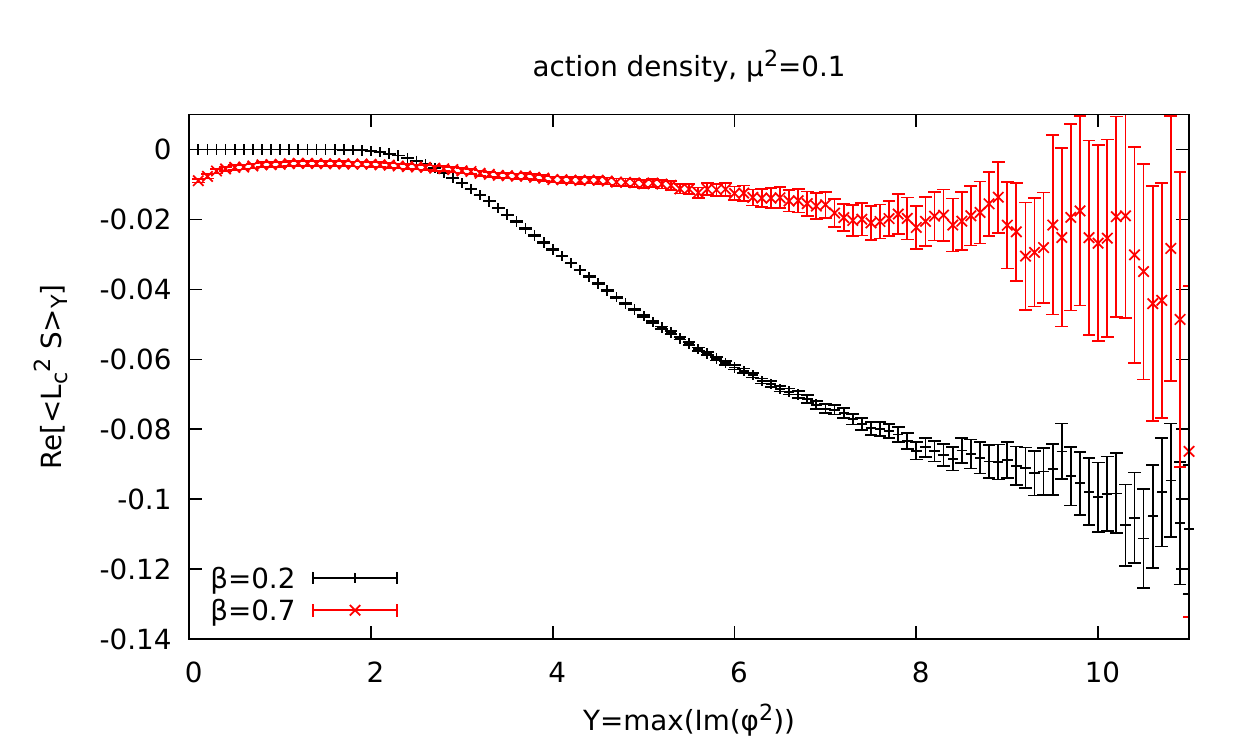} 
\includegraphics[width=0.45\textwidth]{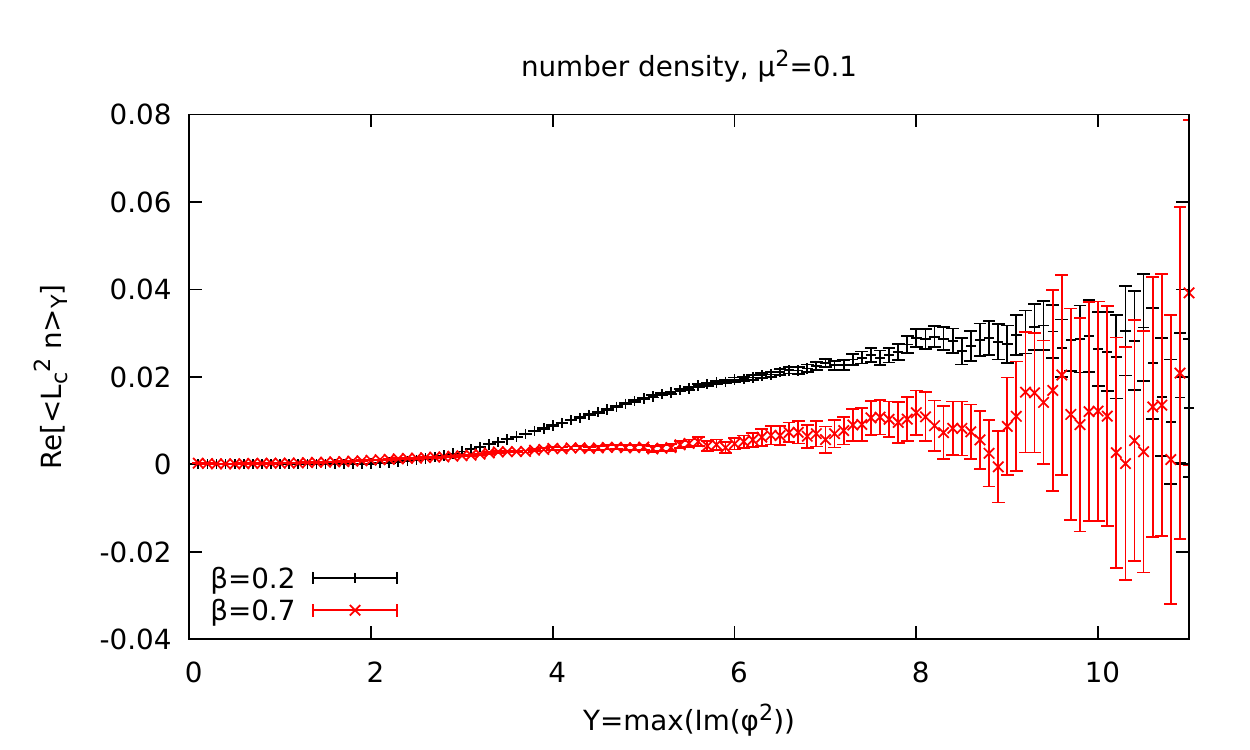} 
\includegraphics[width=0.45\textwidth]{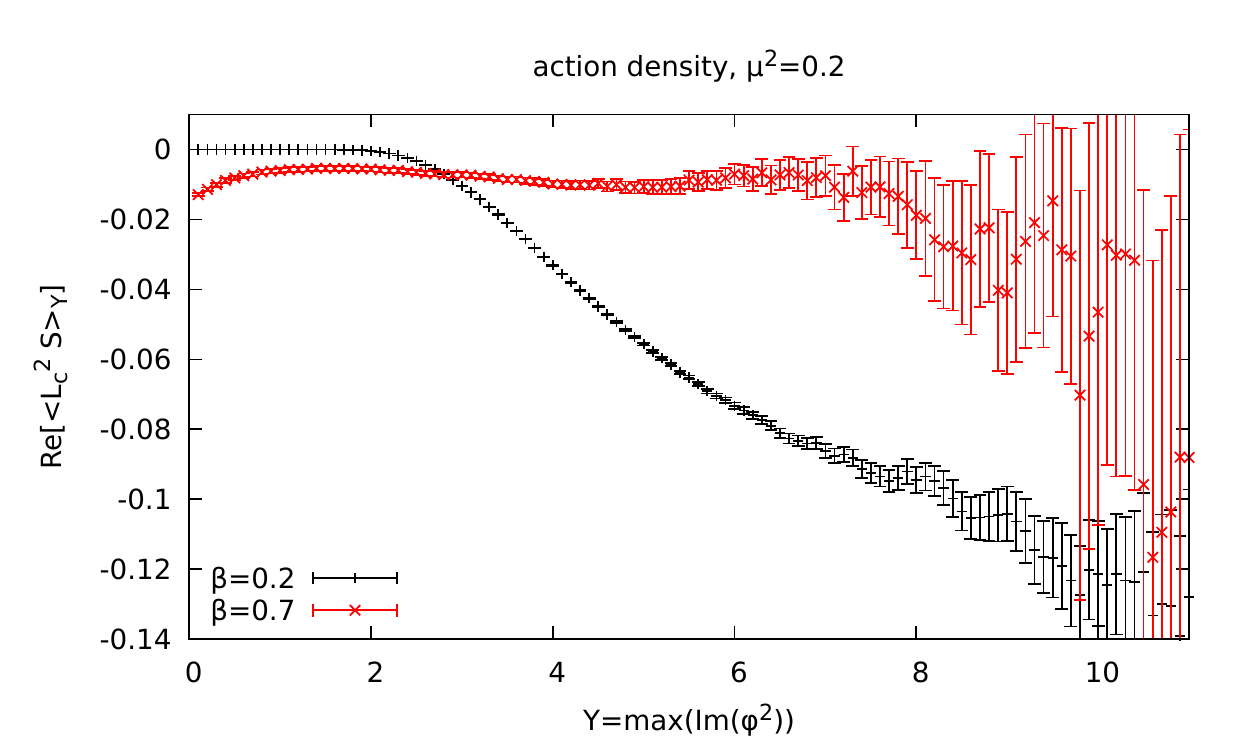} 
\includegraphics[width=0.45\textwidth]{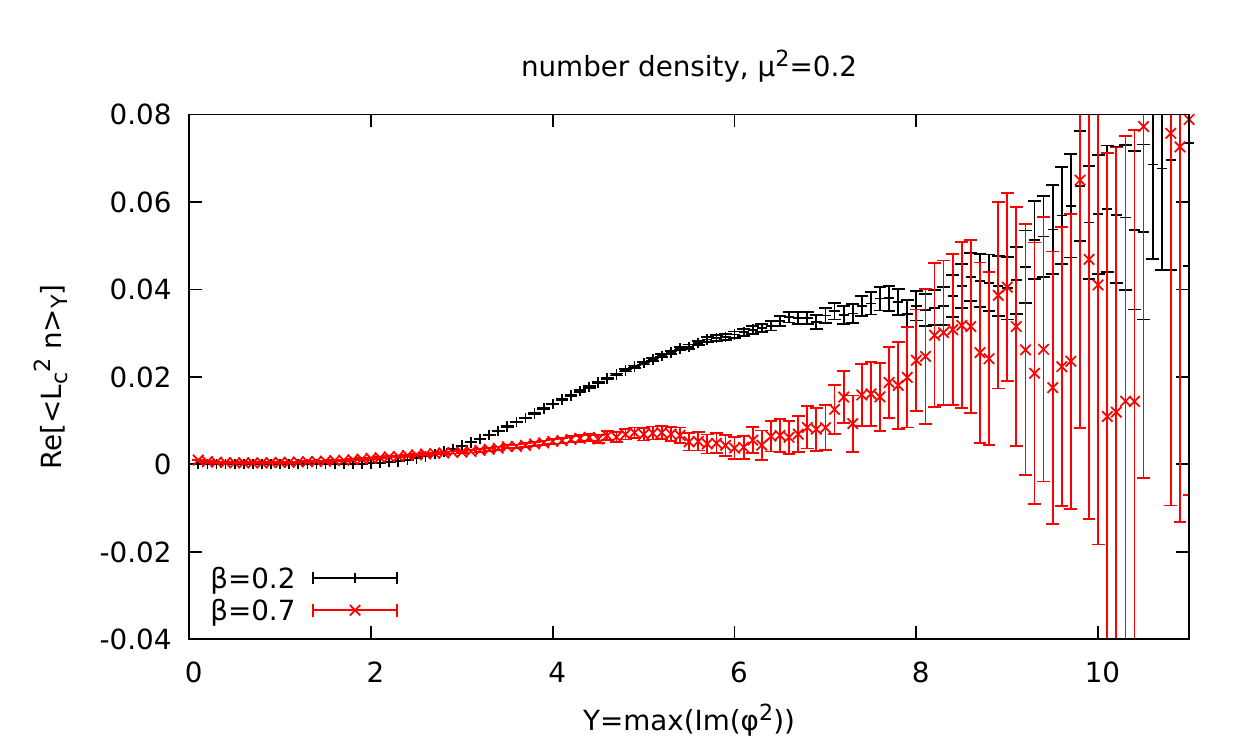} 
\caption{$B_2$ as a function of the cutoff $Y$ in the XY model. Left
  column: $B_2$ for action 
density. Right column: $B_2$ for number density. Top to bottom: 
$\mu^2=10^{-6},0.1,0.2$. } 
\label{fig:XY_BTS2} 
\end{figure}

Next, we analyze the systematic error of the CL method in the XY model.
A straightforward application of $L_c$ yields
the observables for boundary terms $B_2$ of the action density and the number
density, similarly to
the derivation of $B_1$ in Sec.~\ref{sec:XY}, see in App.~\ref{sec:eqappendix}.
We show the resulting $B_2$ in Fig.~\ref{fig:XY_BTS2}, and extract the value of $B_1$ and $B_2$ via fits of a constant to the plateau region in Figs.~\ref{fig:XY_BTS} and \ref{fig:XY_BTS2}, the results are shown in table \ref{XYtable}.
Note that the errors on $B_2$ are rather large 
which is due to the larger
statistical fluctuations of $B_2(Y)$ as well as 
a varying fitting range to the data in Fig.~\ref{fig:XY_BTS2}. Note that there might be an additional systematic error, since $B_1$ reaches its asymptotic value at $Y=10$, while the fitting range for $B_2$ was chosen chosen approximately starting at $Y=7$ up to $Y=11$. Hence, it is possible that $B_2$ has not yet reached its asymptotic value. For $\beta=0.2$ 
  our estimate of the systematic error is close to the measured systematic error of the CL method, where statistically significant deviations can arise due to the lack of a stable plateau region in $B_2$ before the signal becomes too noisy 
  as well as the ansatz (\ref{fttansatz}) not describing $F(t,\tau)$ well enough.
 For $\beta=0.7$ the deviation of CL from worldline is already small, hence a high precision is needed. For this reason we do not investigate $\beta=0.9$ here.

\begin{table}
\begin{tabular}{|l|l|c|c|c|c|c|c|c|}

  \hline
  $\mathcal{O}$ &$\beta, \mu^2$ &  $B_1$ & $B_2 $ & $B_1^2/B_2$ & CL error &CL & worldline & corrected CL\\
  \hline
S&0.2,$10^{-6}$&$0.02567(21)$ & $-0.0730(47)$ & $-0.00902(46)$ & $-0.013029(65)$ & $-0.075316(65)$ & $-0.062288(17)$ & $-0.06630(53)$ \\

  \hline
&0.2,0.1&$0.03309(25)$ & $-0.0903(79)$ & $-0.01213(89)$ & $-0.0169974(91)$ & $-0.0792922(91)$ & $-0.062295(18)$ & $-0.06716(90)$ \\

\hline
&0.2,0.2&$0.03941(28)$ & $-0.109(13)$ & $-0.0142(17)$ & $-0.0205408(80)$ & $-0.0828399(80)$ & $-0.062299(11)$ & $-0.0686(17)$ \\

  \hline\hline
&0.7,$10^{-6}$&$1.440(15)10^{-4}$ & $-7.33(17)10^{-4}$ & $-2.834(46)10^{-5}$ & $-1.23(33)10^{-4}$ & $-1.482311(33)$ & $-1.48219(35)$ & $-1.482283(34)$ \\

   \hline
&0.7,0.1&$0.004783(50)$ & $-0.0082(23)$ & $-0.00278(69)$ & $-0.002791(31)$ & $-1.526766(31)$ & $-1.52398(35)$ & $-1.52399(72)$ \\
   \hline
&0.7,0.2&$0.006013(38)$ & $-0.00873(96)$ & $-0.00414(45)$ & $-0.002488(29)$ & $-1.568899(29)$ & $-1.56641(20)$ & $-1.56476(48)$ \\

  \hline\hline
n&0.2,$10^{-6}$&$4.8(1.6)10^{-5}$ & $-0.00021(124)$ & $1.3(3.7)10^{-5}$ & $1.36(31)10^{-5}$ & $1.36(31)10^{-5}$ & $-1.2(1.1)10^{-8}$ & $0.89(7.65)10^{-6}$ \\

  \hline
&0.2,0.1&$-0.01147(15)$ & $0.0286(32)$ & $0.00460(24)$ & $0.0058177(41)$ & $0.0058182(41)$ & $4.9(2.1)10^{-7}$ & $0.00122(69)$ \\

\hline
&0.2,0.2&$-0.01821(13)$ & $0.047(12)$ & $0.0071(15)$ & $0.0094104(40)$ & $0.0094114(40)$ & $1.04(19)10^{-6}$ & $0.0023(15)$ \\

  \hline\hline
&0.7,$10^{-6}$&$-5.4(1.5)10^{-7}$ & $0.31(1.31)10^{-5}$ & $1.01(79)10^{-7}$ & $-1.15409(62)10^{-4}$ & $4.72951(62)10^{-4}$ & $5.88(82)10^{-4}$ & $4.72849(76)10^{-4}$ \\
   \hline
&0.7,0.1&$-0.00144(6)$ & $0.0031(12)$ & $6.7(1.6)10^{-4}$ & $8.942(52)10^{-4}$ & $0.1557730(52)$ & $0.15488(19)$ & $0.15510(21)$ \\

   \hline
&0.7,0.2&$-0.002501(92)$ & $0.0045(12)$ & $0.00138(20)$ & $0.0010128(59)$ & $0.2280217(59)$ & $0.22701(15)$ & $0.22664(36)$ \\
  \hline
\end{tabular}
\caption{The estimation of the systematic error of CL for the XY model. $B_1$ was extracted by fitting a constant in the range of $Y=10-15$ for $\beta=0.2$ and $Y=5-10$ for $\beta=0.7$ to figure \ref{fig:XY_BTS}. For $B_2$ we chose a fitting range of $Y=8-10$ for $\beta=0.2$ and $Y=2-6$ for $\beta=0.7$ in figure \ref{fig:XY_BTS2}. Errors given are statistical and systematic errors from the fit combined. The systematic error of the fit was estimated by shifting the fitting range by $\pm 1$ and computing the difference in the resulting $B_n$, we choose the maximum value of this deviation as the systematic error estimate for the fit.}
  \label{XYtable}
\end{table}

\subsection{HDQCD}
\label{hdqcderror}

As the numerical estimation of $B_2$ is quite expensive due to large fluctuations and finite stepsize effects, we restricted ourselves here to the calculation of $B_1$ (see also Sec.~\ref{ressec:hdqcd}) which for the spatial plaquette average seems to give an upper bound of the systematic error of CL. A more detailed analysis is delegated to a follow up study including also full QCD.

 In Table.~\ref{HDQCDtable} we show the boundary term for
 the spatial plaquette variable.
 One observes that  the value of $B_1$ is
 roughly a factor of 10 higher than the error of the CL  
 approach, therefore it can be used as an indicator of the 
 magnitude of the systematic error of the CL approach.
 The boundary terms of the Polyakov loop appear much smaller than those
 of the spatial plaquette (consistent with zero inside large statistical errors), in spite of the averages deviating significantly
 from the reweighting result.
 This might signal that the ansatz (\ref{fttansatz}) is too simple
 (and correspondingly the assumption that the maximal slope of $F(t,\tau)$
 is at $\tau=0$ may not be valid)
 for the Polyakov loop observable, or that there are strong stepsize
 effects at play.
 This issue is currently under investigation.

\begin{table}
\begin{tabular}{|c|c|c|c|c|c|c|c|}
  \hline
 $\beta$ &  $B_1$ &  CL error &CL & reweighting \\
  \hline
 5.1 & -0.578(22) & 0.056729(28) &0.471949(27) & 0.4152200(74)\\
 5.5 & -0.2808(99) & 0.020075(24) &0.516855(19) & 0.496780(14)\\
 5.8 & -0.0305(14) & -0.004869(54) &0.566131(53) & 0.5710000(91)\\
 6.0 & -0.00378(49) & $-6.39(25)10^{-4}$ &0.594671(25) & 0.5953100(56)\\
\hline
\end{tabular}
\caption{The boundary terms of the spatial plaquette average 
  in HDQCD on a $6^4$ lattice at $\mu=0.85,\ N_F=1,\ \kappa=0.12$.}
 \label{HDQCDtable}
\end{table}

\section{Conclusions}

We have analyzed the emergence of boundary terms responsible for failure 
of the CL method for various models, from one-plaquette and Polyakov loop 
models to high density QCD (HDQCD) and the XY model.
We used two 
mathematically equivalent versions: `surface' and `volume' and we found 
that numerically they agree wherever both can be computed. The `volume' 
version turns out to be preferable for numerical simulation of HDQCD and 
the XY model. The vanishing/non-vanishing of those terms signals 
correctness/failure of the CL simulations.

  Our analysis should give a quantitative estimate
  for the deviation of the CL method. In practice one must rely on
  a truncated ansatz for the calculation of the interpolating function
  between CL and correct results and thus the numerical costs in some cases
  might be very high.
The drift criterion \cite{Nagata:2016vkn}, on the other 
hand, is easier to use, but it is of a qualitative nature.

 We show that in case the boundary terms are nonzero, one can
  estimate the error of the CL result at the cost of measuring
  a `higher order' boundary term observable.
  The estimation uses an ansatz for the $F(t,\tau)$ function interpolating
  between CL and correct results.
  This allows the calculation of
  the ``corrected CL'' value, which in the case of the $U(1)$ one plaquette
  model gives the correct result to a high accuracy.
  In case of the 3d XY model studied here
  it allows to estimate the size of the systematic error with
  reasonable accuracy.
  In case of HDQCD the boundary term for the spatial plaquette variable
  allows an estimation of the order of magnitude of the systematic error
  that the CL approach has due to nonzero boundary terms.
  A detailed analysis of further observables such as the Polyakov loop average and
  the higher order boundary terms in HDQCD as well
  as full QCD are currently under investigation.

\acknowledgments \noindent

We thank G.~Aarts for illuminating discussions and for direct interest
in our analysis. We thank J.~Nishimura for stimulating discussions. D.~Sexty 
is funded by the Heisenberg programme of the DFG (SE 2466/1-2). 
M.~Scherzer and I.-O.~Stamatescu are supported by the DFG under grant 
STA283/16-2. The authors acknowledge support by the High Performance and 
Cloud Computing Group at the Zentrum f\"ur Datenverarbeitung of the 
University of T\"ubingen, the state of Baden-W\"urttemberg through bwHPC 
and the German Research Foundation (DFG) through grant no INST 37/935-1 
FUGG. Some parts of the numerical calculations were done on the GPU 
cluster at the University of Wuppertal.

\appendix
\section{Higher order boundary terms}
\label{sec:eqappendix}

In the case of the $U(1)$ one plaquette model
defined in eq. (\ref{u1oneplaq}) the observable for $B_2$
is given by
\bea
L_c^2 e^{ikx}  
=& k e^{ikx} \left[ k^3 + 2 ks + 2 i k^2 s x + i s^2 x - k s^2 x^2 +
( 1 + 2 k^2 + s + 2 i k s x ) \beta \sin(x) \right.  \nonumber \\
& \left. + k \beta^2 \sin^2(x) +
\beta \cos(x) ( -2 i k + s x - i \beta \sin(x) ) \right]
\eea

In the XY model, we look at action observable first.
\bea
L_c S=&\left(\nabla+K_x\right) \nabla S(x)=-\nabla K_x -K_x^2,\\
L_c^2 S=&-\left[\nabla_y^2\nabla_x K_x+2(\nabla_yK_x)^2+2K_x\nabla_y^2K_x +K_y\nabla_y\nabla_xK_x+2K_yK_x\nabla_yK_x\right]\\=&-[A_\textrm{XY}+B_\textrm{XY}+C_\textrm{XY}+D_\textrm{XY}+E_\textrm{XY}],
\eea
where we have introduced the notation $A ... E$ for the terms appearing the the last line. For easier readability, we also introduce
the shorthand:
\bea
\phi^{+\nu}=\phi_x-\phi_{x+\hat{\nu}}-i\mu\delta_{\nu,0},\quad\quad
\phi^{-\nu}=\phi_{x-\hat{\nu}}-\phi_x-i\mu\delta_{\nu,0}.
\eea
Using this notation the drift term is written as 
\bea
K_x=&-\nabla S= \beta\sum_{\nu=0}^2 
\left[-\text{sin}(\phi^{+\nu})+\text{sin}(\phi^{-\nu})\right].
\eea
Performing the derivations, one arrives at the following results for the
terms in $ L_c^2 S$:
\bea
A_\textrm{XY}=2\beta\sum_x \sum_\nu[\text{cos}(\phi^{+\nu})+\text{cos}(\phi^{-\nu})]
\eea
\bea
  B_\textrm{XY}=&2\beta^2\sum_x\left[\left(\sum_\nu\left[-\text{cos}(\phi^{+\nu})-\text{cos}(\phi^{-\nu})\right]\right)^2+\sum_\nu\left(\text{cos}(\phi^{+\nu})\right)^2+\sum_\nu\left(\text{cos}(\phi^{-\nu})\right)^2\right]
\eea
\bea
C_\textrm{XY}=-4\beta^2\sum_x\left[\sum_{\nu=0}^2 
  \left[-\text{sin}(\phi^{+\nu})+\text{sin}(\phi^{-\nu})\right]\right]^2
\eea
\bea
D_\textrm{XY}=&\beta^2\sum_x\left[\sum_\nu-\text{sin}(\phi^{+\nu})+\text{sin}(\phi^{-\nu})\right]\left[\sum_\nu\text{sin}(\phi^{+\nu})-\text{sin}(\phi^{-\nu})\right]\\
&-\sum_x\sum_\nu \beta K(x+\nu) \text{sin}(\phi^{+\nu})
+\sum_x\sum_\nu \beta K(x-\nu) \text{sin}(\phi^{-\nu})
\eea

\bea
  E_\textrm{XY}=& -2 \beta K_x^2 
\left[\sum_{\nu}\left[\text{cos}(\phi^{+\nu})+\text{cos}(\phi^{-\nu})\right]\right]
+2\beta\sum_x K_x
\sum_\nu \left( K_{x+\nu} \left[\text{cos}(\phi^{+\nu})\right]
 + K_{x-\nu} \left[\text{cos}(\phi^{-\nu})\right] \right)
\eea
Next we look at the density, given by $ n=\sum_x n_x = i\beta\sum_x\text{sin}(\phi^{+0}) $. For the boundary terms we need
\begin{align} 
 L_c n=&\sum_{x} \left(\nabla_x+K_x\right) \nabla_x n=\nabla^2_x n +K_x\nabla_x n,\\
 L_c^2 n=&\left[\nabla^2_y\nabla^2_x+(\nabla_y^2 K_x)\nabla_x+2(\nabla_yK_x)\nabla_y\nabla_x+K_x\nabla_y^2\nabla_x+K_y\left(\nabla_y\nabla_x^2+(\nabla_yK_x)\nabla_x+K_x\nabla_y\nabla_x\right)\right]n= \\ 
 = & 2 \nabla_x^4 n + 2 \nabla_x n  \nabla_x^2 K_x + 2 \nabla_x^2 n \nabla_x K_x + 2 i \beta^2 \sin(\phi^{+0}) \cos(\phi^{+0})
 + 2 i \beta^2 \sin(\phi^{-0}) \cos(\phi^{-0}) +
 2 K_x \nabla_x^3 n \\ \nonumber
 &+ K_x \nabla_x^3 n + i \beta K_{x+0} \cos(\phi^{+0}) - i \beta K_{x-0} \cos(\phi^{-0}) \\ \nonumber
 &+ \nabla_x n 
\left( K_x \nabla_x K_x  +  \beta \sum_\nu K_{x+\nu}  \cos(\phi^{+\nu})
+ \beta \sum_\nu K_{x-\nu} \cos(\phi^{-\nu}) \right) \\ \nonumber
&+K_x^2 \partial_x^2 n + i \beta K_x K_{x+0} \sin( \phi^{+0} )
 + i \beta K_x K_{x-0} \sin( \phi^{-0} ),
\end{align}
where the derivatives of $n_x$ are given by
\bea 
  \nabla_x n&=i\beta\left( \text{cos} (\phi^{+0})-\text{cos}(\phi^{-0})\right)\\
  \nabla^2_x n&=i\beta\left( -\text{sin} (\phi^{+0})-\text{sin}(\phi^{-0})\right)\\
  \nabla^3_x n&=i\beta\left( -\text{cos} (\phi^{+0})+\text{cos}(\phi^{-0})\right)\\
 \nabla^4_x n&=i\beta\left( \text{sin} (\phi^{+0})+\text{sin}(\phi^{-0})\right).
\eea

\bibliography{mybib}

\end{document}